\documentclass[12pt]{article}
\usepackage{amsmath}
\usepackage{amsthm}
\usepackage{amsfonts}
\usepackage[dvips]{epsfig}
\usepackage{graphicx}
\usepackage{amssymb}
\usepackage{cancel}
\usepackage{pstricks} 
\usepackage{lscape}
\usepackage{enumitem}
\usepackage{cite}

\DeclareFontFamily{U}{mathx}{\hyphenchar\font45}
\DeclareFontShape{U}{mathx}{m}{n}{<-> mathx10}{}
\DeclareSymbolFont{mathx}{U}{mathx}{m}{n}
\DeclareMathAccent{\widebar}{0}{mathx}{"73}

\newcommand{\beq}{\begin{equation}}
\newcommand{\eeq}{\end{equation}}
\newcommand{\barM}{\widebar{M}}

\makeatletter
\newlength{\apb@width}
\newcommand{\autoparbox}[2][c]{\settowidth{\apb@width}{#2}\parbox[#1]{\apb@width}{#2}}

\newcommand{\Cen}[2]{%
  \ifmeasuring@
    #2%
  \else
    \makebox[\ifcase\expandafter #1\maxcolumn@widths\fi]{$\displaystyle#2$}%
  \fi
}
\makeatother

\begin{document}
\vspace{-0.1in}
\begin{flushright}
{\tt KCL-PH-TH/2018-77}, {\tt CERN-TH-2018-273}  \\
{\tt UT-18-29, ACT-05-18, MI-TH-1815} \\
{\tt UMN-TH-3808/18, FTPI-MINN-18/23} \\
\end{flushright}

\vspace{0.05cm}
\begin{center}
{\bf {\large Symmetry Breaking and Reheating after Inflation in \\[0.2cm] No-Scale Flipped SU(5)}}
\end{center}
\vspace{0.05cm}


\begin{center}{
{\bf John~Ellis}$^{a}$,
{\bf Marcos~A.~G.~Garcia}$^{b}$,
{\bf Natsumi Nagata}$^{c}$, \\[0.1cm]
{\bf Dimitri~V.~Nanopoulos}$^{d}$ and
{\bf Keith~A.~Olive}$^{e}$
}
\end{center}

\begin{center}
{\em $^a$Theoretical Particle Physics and Cosmology Group, Department of
  Physics, King's~College~London, London WC2R 2LS, United Kingdom;\\
Theoretical Physics Department, CERN, CH-1211 Geneva 23,
  Switzerland}\\[0.2cm]
  {\em $^b$Physics \& Astronomy Department, Rice University, Houston, TX
 77005, USA}\\[0.2cm] 
  {\em $^c$Department of Physics, University of Tokyo, Bunkyo-ku, Tokyo
 113--0033, Japan}\\[0.2cm] 
{\em $^d$George P. and Cynthia W. Mitchell Institute for Fundamental
 Physics and Astronomy, Texas A\&M University, College Station, TX
 77843, USA;\\ 
 Astroparticle Physics Group, Houston Advanced Research Center (HARC),
 \\ Mitchell Campus, Woodlands, TX 77381, USA;\\ 
Academy of Athens, Division of Natural Sciences,
Athens 10679, Greece}\\[0.2cm]
{\em $^e$William I. Fine Theoretical Physics Institute, School of
 Physics and Astronomy, University of Minnesota, Minneapolis, MN 55455,
 USA}
 
 \end{center}

\vspace{0.1cm}
\centerline{\bf ABSTRACT}
\vspace{0.1cm}

{\small No-scale supergravity and the flipped SU(5)$\times$U(1) gauge group provide an
ambitious prototype string-inspired scenario for physics below the string scale, which can accommodate
the Starobinsky-like inflation favoured by observation when the inflaton is associated with one of the 
singlet fields associated with neutrino mass generation. During inflation, the vacuum remains in the unbroken
GUT phase, and GUT symmetry breaking occurs later when a field with a flat direction (the flaton)
acquires a vacuum expectation value. Inflaton decay and the reheating 
process depend crucially on GUT symmetry breaking, as decay channels open and close,
depending on the value of the flaton vacuum expectation value.
Here, we consider the simultaneous cosmological evolution of both the inflaton and flaton
fields after inflation. We distinguish weak, moderate and strong reheating regimes,
and calculate in each case the entropy produced as all fields settle to their global minima.
These three reheating scenarios differ in the value of a Yukawa 
coupling that introduces mass mixing between the singlets and the {\bf 10}s of SU(5).
The dynamics of the GUT transition has an important impact on the production of gravitinos,
and we also discuss the pattern of neutrino masses we expect in each of the 
three cases. Finally, we use recent CMB limits on neutrino masses to constrain the reheating models,
finding that neutrino masses and the cosmological baryon asymmetry can both
be explained if the reheating is strong.}

\vspace{0.1in}

\begin{flushleft}
December 2018
\end{flushleft}
\medskip
\noindent

\newpage

\section{Introduction}

Inflation~\cite{reviews} is the dominant paradigm for explaining many cosmological puzzles, such as the 
size and age of the Universe, its (approximate) geometrical flatness, homogeneity and
isotropy on large scales and the absence of many unwanted relics from the Big Bang.
Models of inflation commonly postulate a scalar field, the inflaton, whose potential energy
drives inflation and whose quantum fluctuations generate deviations from flatness, 
homogeneity and isotropy. These are thought to have generated the perturbations
measured in the cosmic microwave background (CMB), which are in turn thought to be the
origins of the structures that formed in the Universe subsequently. These perturbations
may be of two types, scalar or tensor. In many models of inflation these perturbations are
expected to be approximately Gaussian and have an almost scale-invariant spectrum.
CMB measurements are consistent with these generic predictions of cosmological
inflation, with the proviso that so far there are only upper limits on the ratio $r$ of tensor
to scalar perturbations~\cite{planck15,Aghanim:2018eyx,planck18}.

The upper limit on $r$ and measurements of the amount of scale-non-invariant tilt $n_s$ in the spectrum
of scalar perturbations exclude many simple models of inflation involving, for example,
potentials that are monomial functions of canonically-normalized scalar fields. However,
the CMB measurements are completely consistent with the earliest model of inflation,
proposed by Starobinsky~\cite{Staro}, in which the standard
Einstein-Hilbert action for general relativity is modified by the addition of an $R^2$ term.
Cosmological perturbations in this model were first calculated by Mukhanov and Chibisov~\cite{MC,Staro2},
and it was shown that the Starobinsky model is equivalent via a conformal transformation~\cite{WhittStelle} to a
theory in which the Einstein-Hilbert action is supplemented by an inflaton field with a potential
that is asymptotically flat at large field values.

One of the most interesting aspects of cosmological inflation and the CMB is that
it offers a window on physics at energy scales far beyond the direct reach of accelerators,
and potentially within a few orders of magnitude of the scale at which quantum-gravitational
effects become important. This has motivated many studies of inflationary models motivated by
theories of quantum gravity, with string theory being foremost among them. One may follow
this line of thought in either of two directions: either top-down - looking for an inflaton
candidate within some specific string model, or bottom-up - taking a phenomenological
approach based on general features expected in effective low-energy field theories
derivable (in principle) from string theory.

In this paper we take the latter approach, studying a model of inflation formulated
in the framework of no-scale supergravity~\cite{no-scale,LN} - which is known to
be the general form of four-dimensional effective field theory derivable from string
theory that embodies low-energy supersymmetry~\cite{Witten}, and assuming
that the sub-Planckian visible-sector gauge symmetry group is flipped 
SU(5)$\times$U(1)~\cite{Barr,DKN,flipped2} - which has been derived in explicit
four-dimensional string models~\cite{AEHN,cehnt}.
A supersymmetric framework is desirable to keep the scale of
inflation naturally small compared with the Planck scale~\cite{cries}, as indicated by the small magnitude $A_s$ of the scalar density fluctuations. We recall also
that the scale of inflation may be comparable with the scale of grand unification, and flipped SU(5) has been shown~\cite{egnno2} to contain
interesting inflaton candidates.

No-scale supergravity initially attracted attention as an interesting framework for
constructing models of inflation~\cite{GL,EENOS} because it yielded naturally a flat potential with no
anti-de Sitter `holes', resolving the so-called $\eta$ problem. Interest in no-scale
inflation was renewed when it was shown~\cite{ENO6,ENO7,EGNO6} to accommodate comfortably values of
the tensor-to-scalar ratio $r$ and the scalar tilt $n_s$ that were very compatible with
the most recent measurements using the Planck satellite~\cite{planck15,Aghanim:2018eyx,planck18},
and potentially very similar to the values predicted~\cite{MC} by the original Starobinsky
model~\cite{Staro}.

This success motivated us to study~\cite{egnno2} the possibility of no-scale inflation within
the framework of flipped SU(5)~\cite{Barr,DKN,flipped2,AEHN,cehnt,eln,Ellis:1992nq,Ellis:1993ks}~\footnote{A similar approach was taken in \cite{EGNNO} in the context of an SO(10) grand unified theory.}.
Flipped SU(5) models of inflation outside the no-scale framework were considered in \cite{fi}.
We showed in~\cite{egnno2} that the available measurements of $n_s$, in particular, imposed
important constraints on model parameters such as the Yukawa couplings, leading also to
significant implications for the model's predictions for neutrino masses
\cite{eln,Ellis:1992nq,Ellis:1993ks}. The model has several
scalar fields, so important topics in~\cite{egnno2} included the identification of the inflaton
and the behaviours of this and other scalar fields during and after inflation. In particular, we
considered various scenarios for baryogenesis, reheating and the GUT phase transition, arguing for 
strong reheating, which would avoid excessive entropy production that might dilute the baryon asymmetry.

We consider in this paper the detailed and coupled evolution of the inflaton and the flaton that is responsible for GUT symmetry breaking. Because of 
finite-temperature corrections to the GUT Higgs potential, the details of GUT symmetry breaking depend 
on reheating, which in turn depends on the strength of 
one of the Yukawa couplings of the inflaton to the 
Higgs and matter. We consider in detail the implications for gravitino production, CMB observables and neutrino masses
of different scenarios for the reheating.

The structure of the paper is as follows. In the next Section, we outline
the basic framework for our no-scale flipped SU(5)$\times$U(1) model, and describe the GUT phase transition in Section \ref{phase}. The bulk of our results are given in Section \ref{sec:2comp},
where we describe the effects of reheating on the evolution of the flaton and its subsequent production of
entropy. We distinguish between strong, moderate and
weak reheating scenarios. Among the consequences of
the different reheating scenarios is the production of 
gravitinos which are produced during reheating and 
subsequently diluted by flaton decay, which is discussed in Section \ref{sec:gravitino}. Constraints on the model
from the CMB and neutrino masses are discussed in Sections \ref{sec:cmb} and \ref{sec:neutrinos} respectively.  Our summary and conclusions are given in Section \ref{sec:sum}.

\section{No-Scale Flipped SU(5)$\times$U(1)}
\label{sec:model}

The no-scale flipped ${\rm SU}(5) \times {\rm U}(1)$ model we consider was described in
detail in \cite{egnno2}, so here we describe only its essential features.

The field content of this flipped ${\rm SU}(5) \times {\rm U}(1)$ GUT consists~\cite{Barr,DKN,flipped2,AEHN} of three generations of Standard Model (SM)
matter fields, each with the addition of a right-handed neutrino and arranged in
a $\mathbf{10}$, $\bar{\mathbf{5}}$, and $\mathbf{1}$ of SU(5). The assignments of the
right-handed leptons, as well as the right-handed up- and down-type
quarks, are ``flipped'' with respect to standard SU(5). The ${\rm SU}(5) \times {\rm U}(1)$ GUT group is
broken to the SM group via
$\mathbf{10}+\overline{\mathbf{10}}$ Higgs representations of SU(5), and
subsequently to the unbroken ${\rm SU}(3) \times {\rm U}(1)$ symmetry via electroweak doublets
in $\mathbf{5}+\bar{\mathbf{5}}$ representations. Our notations for the
fields and their gauge representations are as follows:
\begin{alignat}{3}
&F_i && = ({\bf 10},1)_i  && \ni \; \left\{d^c,Q,\nu^c\right\}_i~,\nonumber\\
&\bar{f}_i &&=(\bar{\bf 5},-3)_i && \ni \; \{u^c,L\}_i~,\nonumber\\
&\ell^c_i &&=({\bf 1},5)_i && \ni \; \{e^c\}_i~,\nonumber\\
&H &&=({\bf 10},1)~,\nonumber\\
&\bar{H} &&=(\overline{\bf 10},-1)~,\nonumber\\
&h &&=({\bf 5},-2)~,\nonumber\\
&\bar{h} &&=(\bar{\bf 5},2) \, ,
\label{eq:charges}
\end{alignat}
where the subscripts $i = 1, 2, 3$ are generation indices that we
suppress for clarity when they are unnecessary. 
The model
also employs four singlet fields, which have no U(1) charges and are
denoted by $\phi_a=({\bf 1},0)$, $a=0,\ldots,3$.  

The superpotential of the theory has the following generic form up to third order in the chiral superfields:
\begin{align} \notag
W &=  \lambda_1^{ij} F_iF_jh + \lambda_2^{ij} F_i\bar{f}_j\bar{h} +
 \lambda_3^{ij}\bar{f}_i\ell^c_j h + \lambda_4 HHh + \lambda_5
 \bar{H}\bar{H}\bar{h}\\ 
&\quad   + \lambda_6^{ia} F_i\bar{H}\phi_a + \lambda_7^a h\bar{h}\phi_a
 + \lambda_8^{abc}\phi_a\phi_b\phi_c + \mu^{ab}\phi_a\phi_b\,, 
\label{Wgen} 
\end{align}
where the indices $i,j$ run over the three fermion families, for
simplicity we have suppressed gauge group indices, and we impose
a $\mathbb{Z}_2$ symmetry  
\beq
H\rightarrow -H \,
\label{eq:z2h}
\eeq
that prevents the mixing of SM matter fields with Higgs colour triplets and
members of the Higgs decuplets. This symmetry also suppresses the
supersymmetric mass term for $H$ and $\bar{H}$, which has the advantage
of suppressing dangerous dimension-five proton decay operators.

The K\"ahler potential for the model is assumed to have the no-scale~\cite{ekn2}
form
\beq
K = -3\ln\left[T+\bar{T}-\frac{1}{3}\left(|\phi_a|^2+|\ell^c|^2 +
f^{\dagger}f + h^{\dagger}h + \bar{h}^{\dagger}\bar{h} + F^{\dagger}F +
H^{\dagger}H+ \bar{H}^{\dagger}\bar{H} \right)\right]\,,
\eeq
where $T$ is the volume modulus. 
Therefore, in the absence of any moduli dependence of the gauge kinetic
function, the scalar potential will have the form 
\beq
V=e^{2K/3}\left(|W_i|^2 + \frac{1}{2}D^aD^a\right)\,,
\eeq
where $W_i$ is the derivative of the superpotential with respect to the $i$th superfield, and the $D$-term part of the potential in the limit of vanishing SM
non-singlets has the form
\begin{equation}
 D^aD^a = \left(\frac{3}{10}g_5^2 +
\frac{1}{80} g_X^2\right)\,\left(|\tilde{\nu}^c_i|^2
+|\tilde{\nu}^c_{H}|^2-|\tilde{\nu}^c_{\bar{H}}|^2\right)^2\,.
\end{equation}
There is a SM singlet that is a linear
combination of $\nu_H^c$ and $\nu_{\bar H}^c$, and is massless in the
supersymmetric limit due to the presence of an $F$- and $D$-flat
direction in the potential.
We denote this combination by $\Phi$, and refer to it as the flaton.
The symmetric minimum at the origin of this flat direction is destabilized by a soft supersymmetry-breaking
mass term, and 
the ${\rm SU}(5) \times {\rm U}(1)$ GUT symmetry is broken
along this direction. The resultant symmetry-breaking pattern
is 
\begin{equation}
 {\rm SU}(5) \times {\rm U}(1) \to {\rm SU}(3)_C \times {\rm SU}(2)_L
  \times {\rm U}(1)_Y ~.
\end{equation}
The flat direction is lifted by a non-renormalizable superpotential term of the form
\begin{equation}
    W_{\text{NR}} = \frac{\lambda}{n! M^{2n-3}_{P}} (H \bar{H})^n ~,
\end{equation}
where $M_P \equiv (8\pi G_N)^{-1/2}$ denotes the reduced Planck mass.
The effective potential for the flaton 
field is 
\begin{equation}
    V_{\text{non-th}} (\Phi) = 
    V_0 -\frac{1}{2}m_{\Phi}^2\Phi^2 + \frac{|\lambda|^2}{[(n-1)!]^2 M_P^{4n-6}} \Phi^{4n-2} ~.
    \label{eq:treeV}
\end{equation}
where $m_\Phi$ denotes the soft mass of $\Phi$. By minimizing this potential, 
we have
\begin{equation}
    \langle \Phi \rangle 
= 
\biggl[
\frac{\{(n-1)!\}^2m_\Phi^2 M_P^{4n-6}}{(4n-2) |\lambda|^2}
\biggr]^{\frac{1}{4(n-1)}} ~.
\end{equation}
Therefore, to obtain a GUT scale vacuum
expectation value (vev) with an ${\cal O}(1)$ $\lambda$, 
we should have $n\geq 4$.  Once the flat direction is lifted, we expect the flaton (and flatino) mass to be 
of order the supersymmetry-breaking scale. For further details, see \cite{egnno2}.

\section{The GUT Phase Transition}
\label{phase}

\begin{figure}[!t]
\centering
    \includegraphics[width=\textwidth]{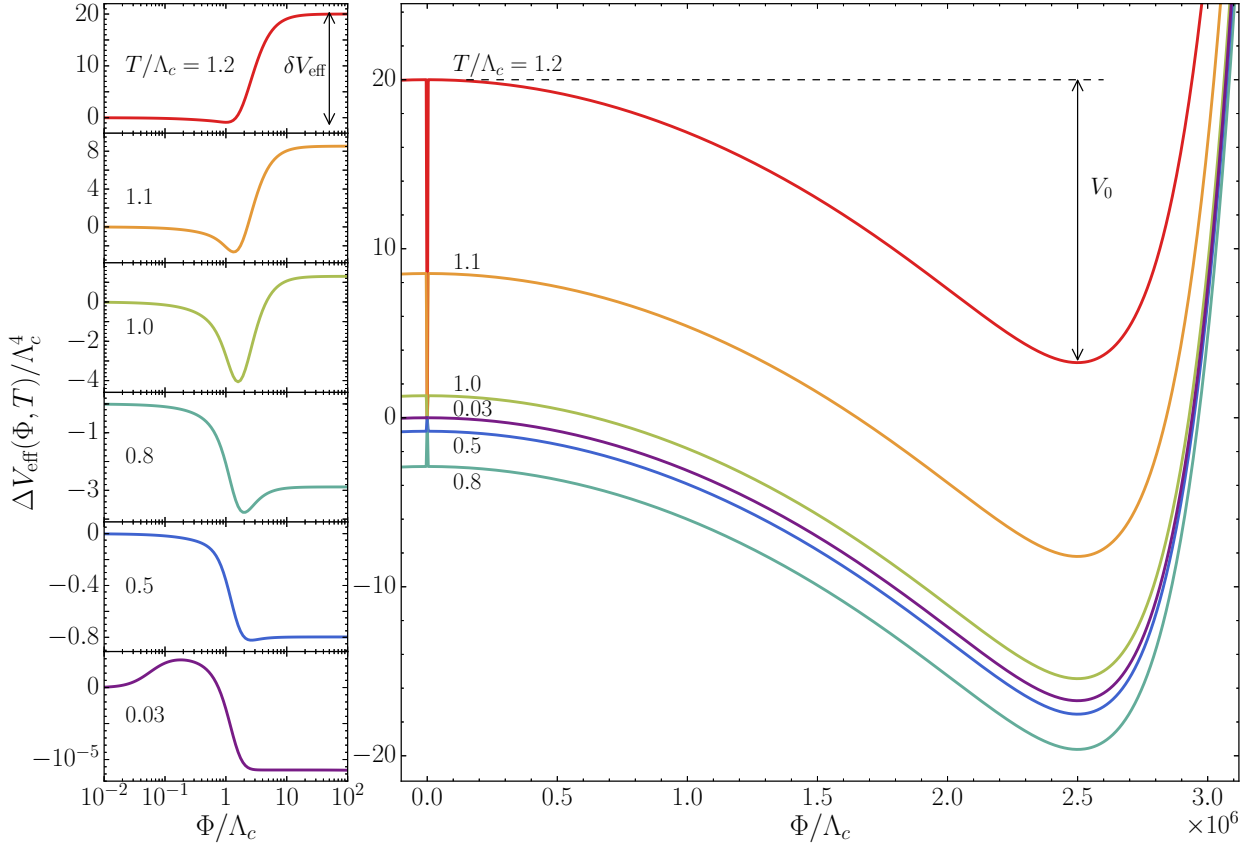}
    \caption{\it The evolution with temperature of the effective potential in strongly-coupled SU(5)$\times$U(1). Here $\Delta V_{\rm eff} = V_{\rm eff}(\Phi,T)-V_{\rm eff}(0,T)$, where $V_{\rm eff}$ includes the non-thermal contribution (\ref{eq:treeV}) with $n=4$, $m_{\rm \Phi}=10\,{\rm TeV}$, $\Lambda_c = 4\times 10^9\,{\rm GeV}$ and $\langle \Phi\rangle= 2.5\times 10^6 \Lambda_c$ at low temperature. The heights of the left and right sides of the barrier $\delta V_{\rm eff}$ and $V_0$ are labelled for the value of $T/\Lambda_c = 1.2$ (see Section \ref{sec:inctr}).}
    \label{fig:veff}
\end{figure}

\begin{figure}[!t]
\centering
    \includegraphics[width=0.8\textwidth]{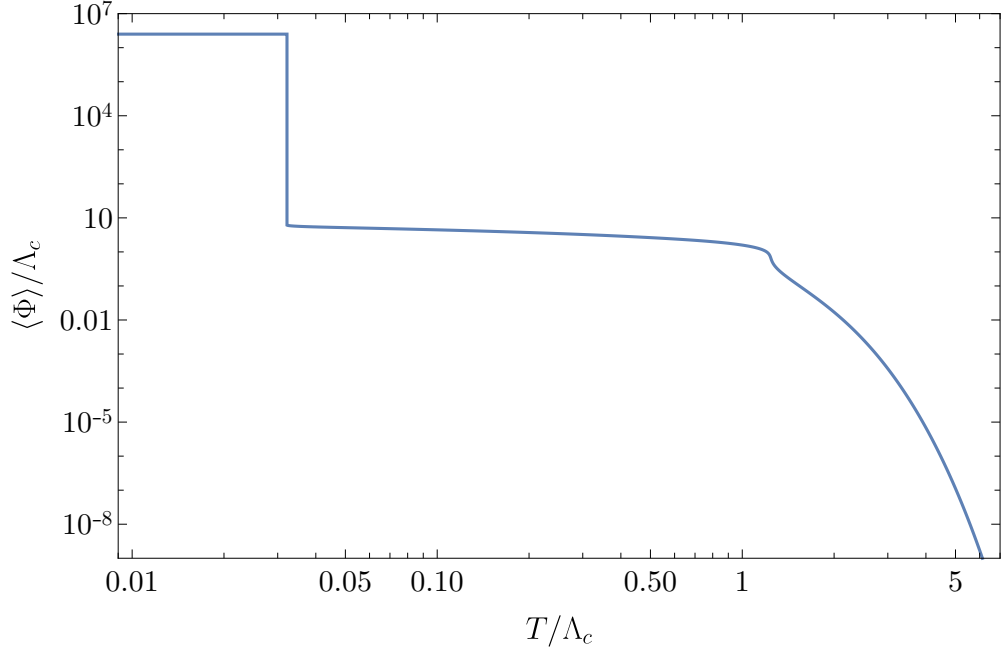}
    \caption{\it The evolution with temperature of $\langle\Phi\rangle$ in strongly-coupled SU(5)$\times$U(1). The parameters are the same as in Fig.~\ref{fig:veff}. }
    \label{fig:vev}
\end{figure}

As discussed in~\cite{supercosm,NOT,Campbell:1987eb,egnno2}, the onset of the SU(5)$\times$U(1) $\rightarrow$ SU(3)$_C\times$SU(2)$_L\times$U(1)$_Y$
GUT symmetry-breaking phase transition is determined by the difference in the number of light degrees of freedom, $g$,
 between the symmetric and Higgs phases of the theory. The light superfields that remain massless in the broken phase 
 contribute to the temperature-dependent correction to the effective potential as follows:
\beq
V_{\rm eff,\,light} = -\frac{\pi^2T^4}{90}g\,.
\eeq
Under the assumption that the chiral and vector couplings that determine the flaton-dependent masses are $\mathcal{O}(1)$ in the strong-coupling domain, a phenomenological fit to the temperature-dependent correction to the effective potential can be written in the form 
\beq\label{eq:Veff}
V_{\rm eff}(\Phi,T) \;\approx\; N_{\Phi}\frac{T^4}{2\pi^2} \sum_{\alpha=0,1}(-1)^{\alpha}\int_0^{\infty}dy\,y^2\ln\left[1-(-1)^{\alpha}\exp\left(-\sqrt{y^2+(\Phi/T)^2}\right)\right]\,,
\eeq 
where $N_{\Phi}$ denotes the number of $\Phi$-dependent massive superfields in the corresponding regime. Figure~\ref{fig:veff} shows the resulting shape of the effective potential as a function of $\Phi$  for $0.03\leq T/\Lambda_c \leq 1.2$. We have used a smooth logistic function to interpolate $g$ and $N_{\Phi}$ around the strong-coupling-transition scale $\Lambda_c$\footnote{The value of $\Lambda_c$ depends on the details of the strong coupling phenomena and has been discussed in some detail in \cite{Campbell:1987eb,egnno2}. Reasonable estimates for $\Lambda_c$ lie between $10^8$ and $10^{14}$ GeV.}. A similar interpolation is used to approximate the integral in (\ref{eq:Veff}). In order to track the evolution of the instantaneous vev of $\Phi$, we have added the thermal correction to the non-thermal effective potential $V_{\text{non-th}} (\Phi)$ in Eq.~\eqref{eq:treeV}.
The location of the local minimum near the origin is shown in Fig.~\ref{fig:vev}. Note that for $1\gtrsim T/\Lambda_c\gtrsim 0.03$, this minimum is metastable, separated from the true vacuum by a shrinking barrier that finally disappears for $T\lesssim 0.03\,\Lambda_c$. 

\noindent

The presence of a barrier separating the metastable minimum from the global minimum at low temperatures suggests that the phase transition could be driven by different mechanisms, depending whether the kinetic energy of the flaton $\Phi$ is dominated by incoherent fluctuations of $\mathcal{O}(T^4)$ as the transition takes place. 

\subsection{Incoherent flaton}\label{sec:inctr}

 If the height of the potential barrier is smaller than the thermal kinetic energy of $\Phi$, the incoherent component of $\Phi$ will drive the transition, destroying any coherent contribution and displacing $\Phi$ to the low-energy vacuum within a Hubble time. As the universe continues to cool down, the energy density of $\Phi$ will simply redshift as radiation. We will refer to this scenario as {\em strong reheating}.
 
\begin{figure}[!t]
\centering
    \includegraphics[width=0.8\textwidth]{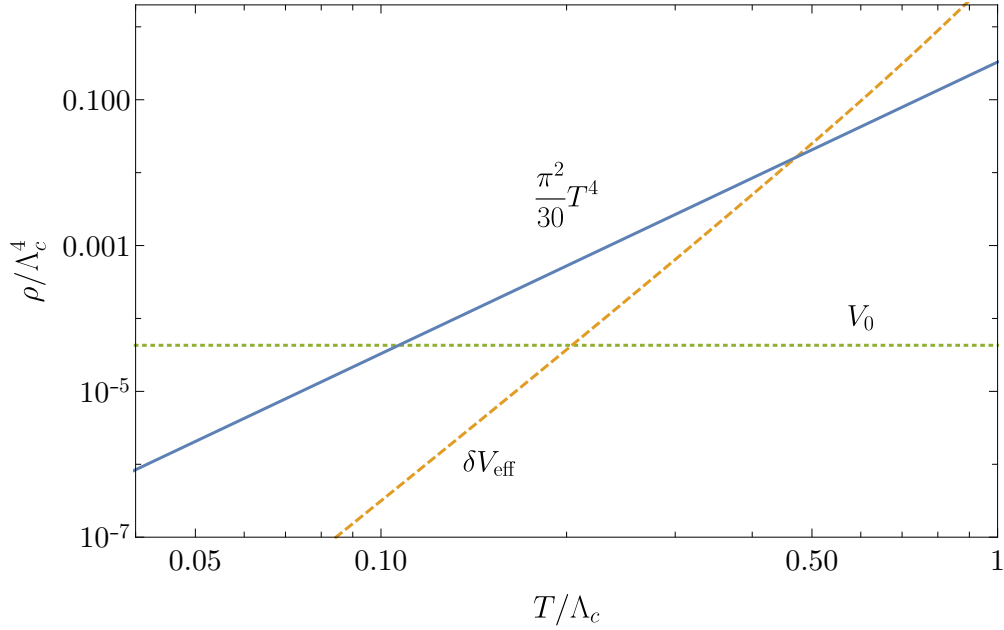}
    \caption{\it The evolution with temperature of the the height of the left side of the barrier $\delta V_{\rm eff}$, the height of the right side of the barrier $V_0$ and the thermal kinetic energy density of the incoherent flaton, in the strong reheating scenario with $\Lambda_c=10\barM$. The conditions (\ref{2cond}) can be seen to be satisfied for $0.47\gtrsim T/\Lambda_c \gtrsim 0.11$.}
    \label{fig:strong_cond}
\end{figure}

 Let us denote by $M_{\rm GUT}\equiv \langle \Phi\rangle_{T\ll \Lambda_c}$ the vev of the flaton in the Standard Model vacuum. 
 Quantitatively, if we denote by $\delta V_{\rm eff}$ the height of the left side of the barrier, $\delta V_{\rm eff}=V_{\rm eff}^{\rm max}-V_{\rm eff}^{\rm meta}$, and we call the height of the right side of the barrier $V_0=(\frac{n-1}{2n-1})m_{\Phi}^2M_{\rm GUT}^2$ (see Fig.~\ref{fig:veff}), the transition will be completed incoherently if/when
 \beq
 \frac{\pi^2}{30}T^4 \geq \delta V_{\rm eff}\quad\text{and}\quad \frac{\pi^2}{30}T^4 \geq V_0\,.
 \label{2cond}
 \eeq
 For $\dot{\Phi}^2 \sim T^4 > \delta V_{\rm eff}, V_0$,
 we expect that within a Hubble time ($\Delta t \sim H^{-1}$), $\Delta \Phi \sim T^2/H \gtrsim M_P$
 and incoherent fluctuations drive the transition.
 The first condition in (\ref{2cond}) can be solved numerically, and leads to
 \beq
 T\lesssim 0.47\,\Lambda_c\,,
 \label{c1}
 \eeq
 while the second condition requires that
 \beq\label{eq:Tstrabove}
 T \;\geq\; \left( \frac{30(n-1)}{\pi^2(2n-1)}\right)^{1/4}\left(m_{\Phi}M_{\rm GUT}\right)^{1/2} \;\approx\; 10^{10}\,{\rm GeV}\,,
 \eeq
(see Fig.~\ref{fig:strong_cond}). These two constraints can be satisfied simultaneously only if \cite{cehno2}
 \beq
 \Lambda_c \;\gtrsim\; 2.8\left(\frac{n-1}{2n-1}\right)^{1/4}\left(m_{\Phi}M_{\rm GUT}\right)^{1/2}\,.
 \label{c2}
 \eeq
For $n=4$, corresponding to the value used in the figures, this constraint
on $\Lambda_c$ becomes, $\Lambda_c > 2.27 \left(m_{\Phi}M_{\rm GUT}\right)^{1/2}=2.27\times 10^{10}\,{\rm GeV}$ \footnote{Our results are not sensitive to the choice of $n$ so long as $n>2$.}, for 
$m_\Phi = 10^4$ GeV and $M_{\rm GUT} = 10^{16}$ GeV. 

{In order to simplify the notation in the following sections, we introduce for convenience the following notation
\beq
\barM\;\equiv\;(m_{\Phi}M_{\rm GUT})^{1/2} \;=\; 10^{10}\,{\rm GeV}
\eeq}
for the geometric mean of the flaton mass and its vev.

\subsection{Coherent flaton}\label{sec:cohdef}

If reheating is not strong, but we still have $T>\Lambda_c$ at some time after inflation, the incoherent component is incapable of driving the transition. However, since the height of the barrier separating the two vacua decreases continuously as $T$ decreases, $\Phi$ could potentially go over the barrier through the accumulation of small thermal fluctuations before it completely disappears, i.e.,~through a first-order phase transition driven by ``thermal tunnelling''. However, since the barrier is very wide, and the temperatures at which the transition would be most favoured are low, $T\ll \Lambda_c$, one can verify numerically that the Euclidean action for the $\mathcal{O}(3)$-invariant bounce solution~\cite{Linde:1981zj}
\beq
S_3 \;=\; 4\pi \int r^2dr\,\left[\frac{1}{2}\left(\frac{d\Phi}{dr}\right)^2 + V_{\rm eff}(\Phi,T)\right]\,,
\eeq 
(where $r^2=\mathbf{x}^2$) is always much larger than the corresponding temperature. As an example, right before the barrier vanishes we obtain $S_3/T\simeq 2\times 10^7$, so that the transition rate is $\Gamma_{\rm T}/H\sim e^{-S_3/T}\ll 1$. Therefore, it is safe to say that when the constraints (\ref{c1}) and (\ref{c2}) for strong reheating
are not satisfied, with $T>\Lambda_c$ initially, the GUT phase transition is of the second order, driven by the classical rollover of $\Phi$ down the potential following the disappearance of the metastable vacuum at $T\sim 0.03\,\Lambda_c$. We refer to this scenario as {\em moderate reheating}.

Note that if $T$ is always smaller than $\Lambda_c$ after inflation, a barrier appears that might trap $\Phi$ near the origin. Its presence would delay or prevent the completion of the phase transition~\footnote{However, the regime $\Phi\ll \Lambda_c$ is the strong-coupling domain of the theory, and our approximations may not be applicable in this case.}. We refer to this low-temperature scenario as {\em weak reheating}.

\begin{figure}[!ht]
\vspace{1cm}
\centering
    \includegraphics[width=0.8\textwidth]{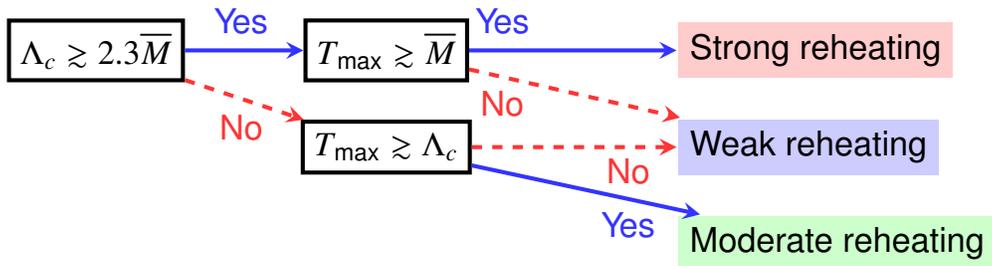}
    \caption{\it Criteria for the classification of the reheating scenarios.}
    \label{fig:classification}
\end{figure}

In summary, the reheating scenarios we consider are classified into three categories: strong, moderate, or weak reheating. The criteria for the classification are summarized in Fig.~\ref{fig:classification}, and we discuss each case in detail in the next Section.

\section{Two-Step Reheating}
\label{sec:2comp}

The inflaton is assumed to be a linear combination of the singlets $\phi_a$, denoted by $S$. 
Specifically, in the basis where $\mu^{ab}$ is diagonal, the inflaton corresponds to the lightest 
state $\phi_0$ with mass $m_s \simeq 3 \times 10^{13}$ GeV, whereas the other three eigenstates 
are assumed to have masses at the GUT scale.\footnote{If we consider a scenario
in which the vev of a singlet field gives rise to the $\mu$-term of $h$ and $\bar{h}$,
then the singlet field becomes the lightest and the inflaton corresponds to
the second lightest state.}
In this case, the vevs of the $\phi_a$ are all {driven rapidly to 0 during inflation}. 
Starobinsky-like inflation occurs when $\mu^{00} = m_s/2$ and $\lambda_8^{000} = - m_s/(3\sqrt{3}M_P)$ along the real direction of $S$,
which we denote as $s$, where $m_s$ denotes the inflaton mass.
In \cite{egnno2} we discussed two possibilities for the inflaton-neutrino coupling.
If $\lambda_6^{i0} = \lambda_7^i = \mu^{0i} = 0$ (which is true when $R$-parity is exact), the fermionic component of $S$ does not mix
with neutrinos. 
Alternatively, it is possible that one linear combination of the $\phi_a$ (denoted by $\phi_{a'}$) acquires a vev.
In this case, $\lambda_7^{a'}$ provides a $\mu$ term for the weak-scale Higgs doublets. If $\lambda_6^{ia'} = 0$, $R$-parity violation is suppressed. However, if $\lambda_6^{i0} \ne 0$, the inflaton will couple to $F_i$. We return to the question of neutrino masses in Section~\ref{sec:neutrinos} but, for now, we restrict our discussion to the latter scenario (called B in \cite{egnno2}). 

The coupling of the inflaton  to light degrees of freedom is dependent on the degree of GUT symmetry breaking. As we discuss in Section~\ref{sec:decays}, for $\langle \Phi\rangle<m_s$, the inflaton decays primarily to various components of the matter and Higgs 10-plets $F_i$ and $H$. Since many of these final states become kinematically forbidden for $\langle \Phi\rangle>m_s$, the decay is completed through the $\nu^c_i~\Phi$ channel. Note, however, that the magnitude of the flaton vev is determined by the temperature of the plasma of decay products, which is in turn determined by the decay rate {and} by the number of light degrees of freedom present in the thermal bath. This circularity of the analysis requires a careful account of the reheating process, which is presented in Section~\ref{sec:reheating}.

\subsection{Inflaton decay channels}\label{sec:decays}

\subsubsection{Inflaton decay with a non-vanishing flaton vev}

We studied in~\cite{egnno2} the decay of the inflaton into leptons of the first generation~\footnote{Decays
into either the second or third generation are constrained by a combination of the reheating
temperature, which constrains the coupling $\lambda_2^{ij} \sin \theta$,
and the sum of neutrino masses. It was found in~\cite{egnno2} that these constraints are most relaxed when $i,j = 1$ (in a diagonal basis).}
in the presence of a large flaton vev, $\langle \Phi\rangle>m_s$. This decay channel arises from neutrino-inflaton mixing, and proceeds with the rate
\beq\label{eq:decnvan}
\Gamma(s\rightarrow L_j\tilde{h}_u)=\Gamma(s\rightarrow \tilde{L}_jh_u)\;=\; \frac{|\lambda_2^{1j}\sin\theta|^2}{8\pi} \, m_{N_{12}}\,,
\eeq
where the mixing angle is given by
\beq
\tan 2\theta\;=\; -\frac{2\lambda_6^{10}\langle \Phi\rangle}{m_s}\,,
\eeq
and the eigenmasses of the heavier states related to $\nu^c_1$ and the fermionic partner of $S$, ${\tilde S}$, can be written as
\beq
m_{N_{11,2}} \;=\;\frac{1}{2}\left[m_s\mp\sqrt{\left(2\lambda_6^{10}\langle\Phi\rangle \right)^2 + m_s^2}\right]\,.
\label{hnumass}
\eeq
If we further assume that $|\lambda_6^{10}\langle\Phi\rangle|\ll m_s$, as we justify below, the decay rate can be rewritten as
\beq\label{eq:stoLh}
\Gamma(s\rightarrow L_j \tilde{h}_u) \;=\; \frac{|\lambda_2^{1j}|^2}{8\pi} \left|\dfrac{\lambda_6^{10}\langle\Phi\rangle}{m_s}\right|^2\, m_s \,,
\eeq
or, in terms of the effective Yukawa coupling $y=(8\pi \Gamma_s/m_s)^{1/2}$,
\beq
y_2 \;=\; |\lambda_2^{1j}| \left|\dfrac{\lambda_6^{10}\langle\Phi\rangle}{m_s}\right|  \;\simeq\; 10^{-5} \left(\frac{M_{\rm GUT}}{m_s}\right) |\lambda_6^{10}| \;\simeq\; 3\times 10^{-3}\, |\lambda_6^{10}|\,.
\label{y2}
\eeq
Here we have made use of the fact that $\lambda_2^{11} \simeq m_u/\langle \bar{h}^0\rangle \simeq 10^{-5}$. The subscript on $y$ 
indicates the dependence of the decay rate on $\lambda_2$.

\subsubsection{Inflaton decay with vanishing flaton vev}\label{sec:novev}

When $\langle \Phi\rangle < m_s$, the inflaton decays to $F$ and $\bar{H}$, with a rate given by
\beq\label{eq:decvan}
\Gamma(s\rightarrow F_i\bar{H}) \;\simeq\; 10\times \frac{|\lambda_6^{i0}|^2}{8\pi} \left( 1- \frac{\langle \Phi\rangle^2}{m_s^2}\right) m_s\,.
\eeq
 The effective Yukawa coupling in this case  is
\beq
y_6^0 \;=\; \sqrt{10}\,|\lambda_6^{i0}| \, ,
\label{y6}
\eeq
where the superscript on $y_6$ refers to the case of a small vev for $\Phi$, $\langle \Phi \rangle \ll m_s$.
However, many of the final-state decay channels disappear if/when $\langle \Phi\rangle$ becomes larger than $m_s$,
since the fields in ${\bar H}$ obtain masses $\propto \langle \Phi \rangle$, and these final states becomes 
kinematically forbidden. Nevertheless, one final state remains open,
namely $\nu^c_i ~ \Phi$, since we see from Eq. (\ref{hnumass}) that one of the
heavy neutrino eigenstates has a mass less than $m_s$. For $\lambda_6^{10} \langle \Phi \rangle \ll m_s$, this state is almost purely $\nu^c_i$ in $F_i$ and $\Phi = (\nu^c_H + \nu^c_{\bar{H}} )/\sqrt{2}$,
with mass much lighter than $m_s$.
When the kinetic suppression factor in Eq. (\ref{eq:decvan}) is absent
the inflaton decay rate becomes
\beq
\Gamma(s\rightarrow \nu^c_{u_i} \Phi) \;\simeq\;  \frac{|\lambda_6^{i0}|^2}{16\pi} m_s \, .
\label{Glate}
\eeq
Thus, as the symmetry is broken the inflaton decay rate drops by a factor of $\simeq$ 20,
and once the symmetry is broken, the effective Yukawa coupling is
\beq
y_6^\Phi \;=\; \sqrt{\frac12}\,|\lambda_6^{i0}| \, ,
\label{y6late}
\eeq
where now the superscript on $y_6$ refers to a GUT scale vev for $\Phi$.
Comparing the couplings in Eqs. (\ref{y2}) and (\ref{y6late}),
we see that even the late decays are dominated by the single channel in 
the $F \bar{H}$ final state and decays into $L h$ can safely be ignored. \footnote{An exception to this conclusion occurs
when $\lambda_6^{10} \gg m_s/2\langle \Phi \rangle$. In that case,
the mass of $\nu^c_i$ is also larger than $m_s$, and all ten $F_i  \bar{H}$ channels
are forbidden after symmetry breaking.
However, in that case, as we discuss below, the decay of the
inflaton is complete before the GUT transition occurs.}

\subsection{Reheating}\label{sec:reheating}

We consider next the evolutions of the energy densities of the inflaton and its decay products, the temperature of the latter, 
as well as the evolution of $\Phi$. As the $F\bar{H}$
decay rate (\ref{eq:decvan}) is in principle sensitive to the Yukawa couplings 
for all three lepton generations, for definiteness we focus only on the case $i=1$. In order for the analysis to be complete it
is necessary to note that, immediately after inflation, the Universe starts in a super-cooled state, $T\rightarrow 0$. 
Assuming that the decay products of the inflaton {\em thermalize instantaneously}, their instantaneous temperature grows
rapidly  as the inflaton $s$ begins oscillating and decaying, until it reaches its maximum value $T_{\rm max}$, 
after which it decreases to $T_{\rm reh}$ and below \cite{ckr,Giudice:1999am,EGNOP}. 
In \cite{egnno2}, we assumed that inflaton decay was dominated by the
decay channel with effective Yukawa coupling $y_2$ in Eq.~(\ref{y2}). However, during inflation and,
more importantly, when exponential inflationary expansion
ends, $\langle \Phi \rangle = 0$, and $y_2 = 0$. As the fields evolve, and the flaton picks up its vev, this channel opens up again, though it does not dominate the decay.\footnote{If we instead couple the inflaton field to the third generation ($i=3$), the $y_2$ decay channel will dominate the decay after the phase transition; in this case, $y_2 \simeq y_t M_{\rm GUT} |\lambda_{6}^{30}|/m_s$, which is much larger than $y_6^\Phi = |\lambda_{6}^{30}| / \sqrt{2}$. However, as we will see,
if we couple the inflaton to the third generation, light neutrino masses will be too
large unless $\lambda_6 \gtrsim 0.01$. But in this case, the reheating is complete before the GUT transition occurs, negating the utility of the potentially stronger coupling $y_2$.}
In order to ascertain the history of the reheating process, we must track the dynamics of the flaton as the temperature grows past $\Lambda_c$,
and then as it decreases below it, simultaneously with the evolution of the inflaton.

\subsubsection{Increasing temperature}\label{sec:inctemp}

At the end of inflation, as the first oscillations of the inflaton begin to decay, the temperature of the radiation produced by inflaton decay
rises to a maximum temperature. 
For definiteness, we make a series of simplifying assumptions. The first assumption is that of a
discontinuous inflation $\rightarrow$ reheating transition. Disregarding for now the coupling of the inflaton to the other singlets $\phi_i$ 
(which is justified for $\lambda_8^{00i}=\lambda_8^{0ij}=0$), the equation of motion for $s$ can the be written as
\begin{flalign}\label{eq:inflations}
& & \Cen{3}{\ddot{s} + 3H\dot{s} + \partial_sV \;=\; 0 \,}      &&  
\end{flalign}
during inflation. Immediately after inflation ends, when the acceleration of the cosmological scale factor, $\ddot{a}=0$ or $\dot{s}^2=V(s)$, we write
\begin{flalign}\label{eq:reheatings}
&  & \Cen{3}{\ddot{s} + (3H + \Gamma_s)\dot{s} + \partial_sV \;=\; 0 \,}      &&  
\end{flalign}
during reheating, i.e.,~we assume that the decay rate turns on instantaneously~\footnote{In practice, this is a safe assumption that is almost always made in the literature,
since any decays occurring before the end of inflation are redshifted away by the continuing accelerated expansion.}. 
Next we approximate the energy density of $s$ by its average value during oscillations. 
With $\langle \rho_s\rangle = \langle \dot{s}^2/2\rangle + \langle V\rangle \simeq \langle \dot{s}^2\rangle$, 
we can average (\ref{eq:reheatings}) to obtain the system of equations that determines the transfer of energy from the inflaton $s$ to its decay products $\gamma$:
\begin{align} \label{eq:rhophieq}
\dot{\rho}_{s} + 3H\rho_s &= -  \Gamma_s\rho_s\,,\\ \label{eq:rhogammaeq}
\dot{\rho}_{\gamma} + 4H\rho_{\gamma} &= \Gamma_s\rho_s\,,\\ \label{eq:hubbleeq}
\rho_s + \rho_{\gamma} &= 3M_P^2H^2\,.
\end{align}
For fixed $\Gamma_s$ this system of equations can be solved formally~\cite{EGNO5}:
\begin{align}
\rho_{s} \;&=\; \rho_{\rm end} \left(\frac{a(v)}{a_{\rm end}}\right)^{-3} e^{-v}\,,\\
\rho_{\gamma}\;&=\; \rho_{\rm end} \left(\frac{a(v)}{a_{\rm end}}\right)^{-4} \int_0^v \left(\frac{a(v')}{a_{\rm end}}\right) e^{-v'}\,dv'\,,
\end{align}
where
\beq
v\;\equiv\;\Gamma_s(t-t_{\rm end}) \, .
\eeq
In terms of the parameter
\beq
A \;=\; \frac{\Gamma_s}{m_s}\left(\frac{3}{4}\frac{\rho_{\rm end}}{m_s^2M_P^2}\right)^{-1/2}\,,
\eeq
the energy density of the relativistic decay products at very early times, $v\ll 1$, can be approximated by~\cite{EGNOP}
\beq\label{eq:rhogammaapp}
\rho_{\gamma} \;=\; \rho_{\rm end} \left(\frac{v}{A}+1\right)^{-8/3} \int_0^{v} \left(\frac{v'}{A}+1\right)^{2/3} e^{-v'}\,dv'\,.
\eeq
The maximum of $\rho_{\gamma}$, and therefore of $T$, is found for
\beq
v_{\rm max} \simeq 0.80A\,,\qquad T_{\rm max} \simeq 0.74 \left(\frac{\Gamma_s m_s M_P^2}{g_{\rm max}}\right)^{1/4}\,.
\eeq
Alternatively, in units of the inflaton mass, $T_{\rm max}$ is reached when $m_s(t-t_{\rm end}) = 2.21$,
i.e., within the first oscillation of the inflaton, independently from the decay rate $\Gamma_s$. 
At even earlier times, $v\ll A$, Eq.~(\ref{eq:rhogammaapp}) predicts the following scaling of the temperature with time:
\beq\label{eq:Tup}
T \simeq \left(\frac{30\rho_{\rm end}}{\pi^2 g} \right)^{1/4} v^{1/4}\,.
\eeq\par
During inflation, the flaton vev is kept at zero due to the large induced mass, $m_{\Phi}^2 \sim m_s^2 e^{\sqrt{2/3}\,s/M_P}$. 
Therefore, as reheating begins, $\Phi=0$ and the inflaton starts decaying through the $F \bar{H}$ channel, 
governed by $\lambda_6$ (or the effective Yukawa coupling, $y_6^0$, given by Eq.~(\ref{y6})). 
For $T\gtrsim \Lambda_c$, the form of the effective potential ensures that this is the dominant decay channel. 
However, for $0< T \lesssim \Lambda_c$ the effective potential favors $\Phi\sim M_{\rm GUT}$. 
Therefore, if $\Phi$ can roll~\footnote{At this stage thermal effects are not expected to lead to a loss of coherence for the flaton.} 
a significant distance away from the origin before $T\sim \Lambda_c$, the GUT symmetry may be broken, 
shutting off many of the $F \bar{H}$ final states and leading to a decay controlled by the $\nu^c_i ~ \Phi$ final state (with $y_6^\Phi$ given by Eq. (\ref{y6late})), at least for a while until $T$ grows past $\Lambda_c$. In units of $\Gamma_s^{-1}$, the amount of time it takes to reach this critical temperature is given by 
\beq
v_c \simeq \left(\frac{\pi^2 g}{30 \rho_{\rm end}}\right) \Lambda_c^4\,,
\eeq
according to (\ref{eq:Tup}). With $g=1545/4$ in the unbroken phase and $\rho_{\rm end}\simeq 0.175 m_s^2M_P^2$ 
for pure Starobinsky inflation~\cite{EGNO5}, $v_c \simeq 5\times 10^{-23}$, which implies that the 
broken phase is not realized before $T_{\rm max}$. This time scale is much smaller than that determined by the Hubble parameter, since $H\sim m_s$:
\beq
m_s(t_c-t_{\rm end}) \simeq 10^{-21} y^{-2}\,.
\eeq
Moreover, the induced thermal mass for $\Phi$ is too small to allow for a fast roll of the field, $H\gg T$.\\

\begin{figure}[ht!]
\centering
    \includegraphics[width=0.9\textwidth]{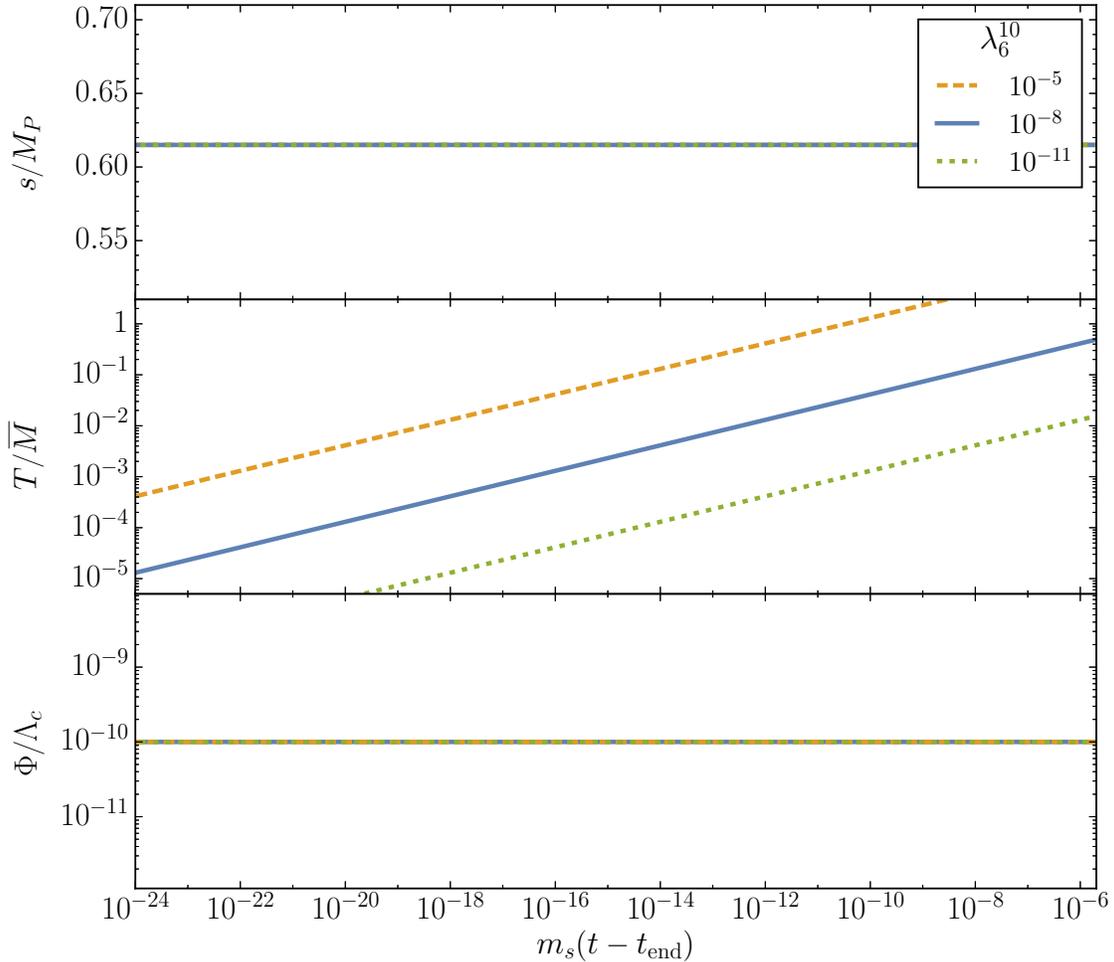}
    \caption{\it Time dependence of the inflaton $s$, the temperature $T$ and the flaton $\Phi$ during the earliest stages of reheating for selected values of $\lambda_6^{10}$. The time-dependence is identical for $\Lambda_c=10\,\barM$ (strong reheating) and $\Lambda_c=0.4\,\barM$ (moderate, weak reheating).}
    \label{fig:reh_early}
\end{figure}

Fig.~\ref{fig:reh_early} shows the result of integrating Eqs.~(\ref{eq:reheatings}), (\ref{eq:rhogammaeq}) and (\ref{eq:hubbleeq})
numerically for the three different values of the Yukawa coupling $\lambda_6^{10}=\{10^{-5}, 10^{-8},10^{-11}\}$. {For $\Lambda_c=10\,\barM$ (strong, weak reheating) and $\Lambda_c=0.4\,\barM$ (moderate, weak reheating) we have verified that the evolution is identical}. The classical roll of the flaton is calculated by integrating numerically the equation
\beq\label{eq:flatonclass}
\ddot{\Phi} + 3H\dot{\Phi} + \partial_{\Phi}V_{\rm eff}(\Phi,T) \;=\;0\,,
\eeq
with the effective potential given by a logistic interpolation of (\ref{eq:Veff}).
 The instantaneous temperature corresponds to
\beq
T \;=\; \left(\frac{30\rho_{\gamma}}{\pi^2 g}\right)^{1/4}\,,
\eeq
where the effective number of relativistic degrees of freedom depends on the masses of the fields present in the thermal bath, 
which are in turn determined only by the vev of the flaton, assuming $T$ is above the EW phase transition. 
For simplicity, we have chosen the parametrization
\beq \label{eq:dofub}
g \;=\; \begin{cases}
1545/4\,, & \Phi<T\,,\\
915/4 \,, & \Phi>T\,,
\end{cases}
\eeq
for the number of degrees of freedom in the unbroken and broken GUT phases.
In all cases we observe a static flaton for $T<\Lambda_c$, confirming that the SU(5)$\times$U(1) symmetry is unbroken for $v<v_{\rm max}$. 
In order to explore the possible roll of the flaton we have assumed the non-vanishing initial condition $\Phi/\Lambda_c=10^{-10}$, though our results would be similar for other choices of $\Phi/\Lambda_c \lesssim 1$.

\begin{figure}[!t]
\centering
    \includegraphics[width=0.9\textwidth]{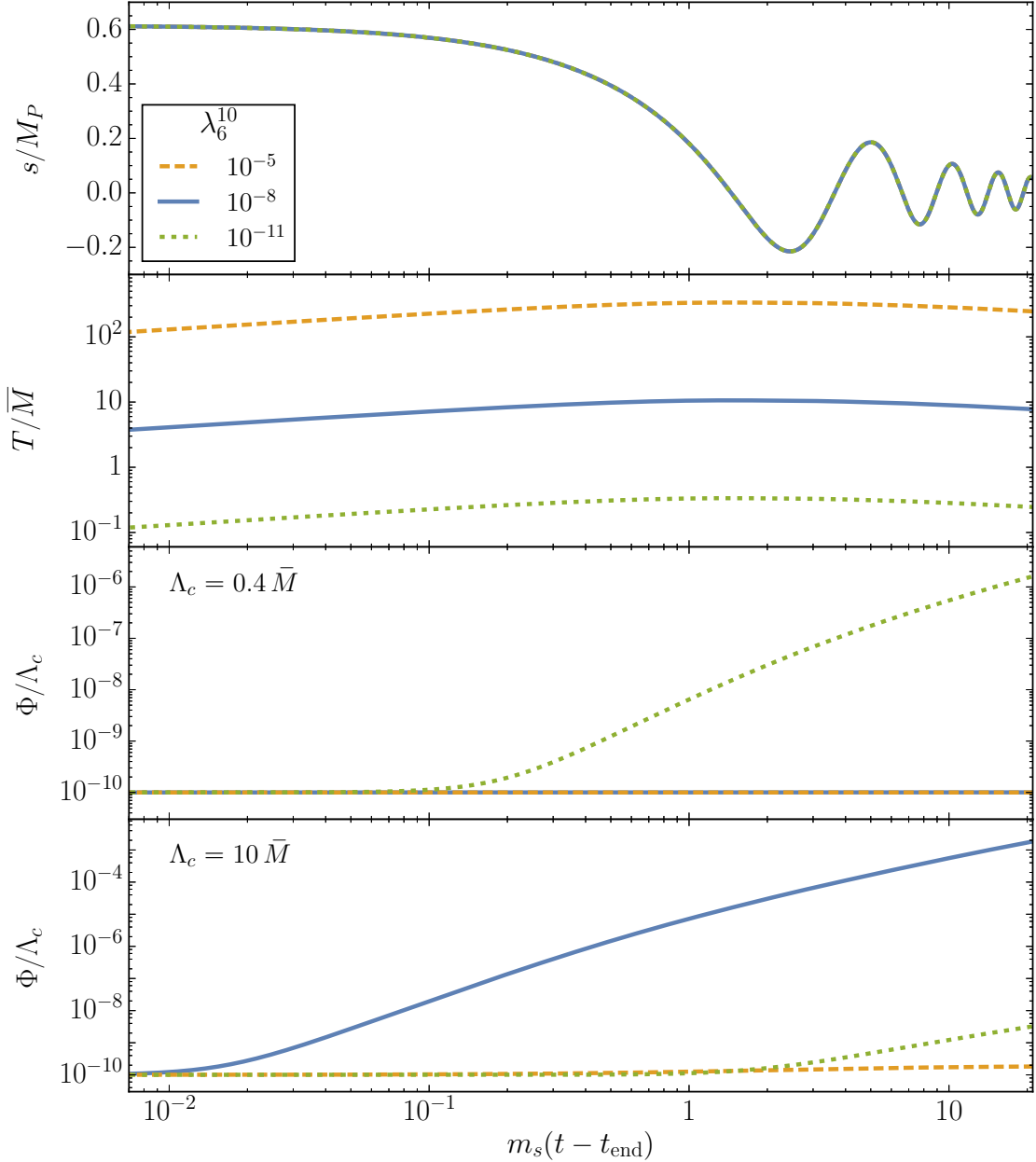}
    \caption{\it Time dependence of the inflaton $s$, the temperature $T$ and the flaton $\Phi$ around $T\sim T_{\rm max}$ for the same values of $\lambda_6^{10}$ and $\Lambda_c$ as Fig.~\ref{fig:reh_early}. }
    \label{fig:reh_osc}
\end{figure}

Fig.~\ref{fig:reh_osc} shows the dynamics of the inflaton, the temperature and the flaton at slightly later times when $T\sim T_{\rm max}$ for the same values of $\lambda_6^{10}$ and $\Lambda_c$ as in Fig.~\ref{fig:reh_early}. 
The first and second panels show that, for the two largest values of $\lambda_6^{10}$, temperatures larger than $\Lambda_c$ can be reached well before the first oscillation of the inflaton $s$. For $\lambda_6^{10}=10^{-11}$, the temperature is comparable to $\Lambda_c$ only close to $v_{\rm max}$ (defined by the the value of $v$ when $T = T_{\rm max}$) in the moderate (weak) reheating case when $\Lambda_c = 0.4 \barM$; for $\Lambda_c = 10 \barM$, $T_{\rm max}\ll \Lambda_c$.
The third and fourth panel show that the flaton remains at in the unbroken minimum beyond $v_{\rm max}$, for our two nominal choices of $\Lambda_c$.

Before we continue with the evolution of the fields beyond $v_{\rm max}$, let us bear in mind two assumptions we have made that could in principle affect
our results:
\begin{enumerate}
\item Our results assume that $\Gamma_s$ turns on instantaneously after inflation. Some quantities may depend on the validity of this assumption (the value and location in time of $T_{\rm max}$, for example).  However, as noted earlier, any decay products generated before $t_{\rm end}$ would redshift away quickly
and we do not expect this assumption to have a large effect on our results.

\item The assumption of instantaneous thermalization is most likely incorrect. 
While thermalization is rapid on the time scale associated with $T_{\rm reh}$ \cite{Davidson:2000er,Harigaya:2013vwa,EGNOP},
at the very early times prior to $v_{\rm max}$, this assumption can be questionable \cite{ag}.
Note that the gauge coupling constant $\alpha_c \simeq 0.0663 \ll 1$ at the time of formation of bound states. This delay in thermalization may allow the flaton to evolve away from the unbroken phase, as the maximum temperature would now be reached at later times. Of course, one would then have to  
consider the non-thermal correction to the effective potential due to in-medium effects.
We note that large Yukawa couplings would 
hasten thermalization and thus our approximation may well be justified.

\end{enumerate}\par\bigskip

After the maximum temperature $T_{\rm max}$ is reached, the combination of the redshift during expansion and the continuous replenishing of relativistic particles into the plasma from $s$ decay lead to an overall decrease of $T$. For $A\ll v\ll 1$ \cite{EGNOP,gmop},
\beq\label{eq:Tlatereh}
T\;\simeq\; \left(\frac{24}{\pi^2 g}\right)^{1/4} (\Gamma_s M_P)^{1/2} v^{-1/4}  \,.
\eeq
Unlike earlier times with increasing $T$, the time dependence of the temperature is gentle, potentially allowing for the phase transition to occur before the end of reheating, where
\begin{align}\notag
\label{eq:Trehstr}
T_{\rm reh} & \;=\; \left(\frac{40}{\pi^2 g_{\rm reh}}\right)^{1/4}(\Gamma_sM_P)^{1/2} \;=\; \left(\frac{5}{8\pi^4 g_{\rm reh}}\right)^{1/4}(m_s M_P)^{1/2} y \\
& \;\approx\;  5.4 \times 10^{14} {\rm GeV}\, \left(\frac{m_s}{3 \times 10^{13} {\rm GeV}}\right)^{1/2} \left(\frac{1545/4}{g_{\rm reh}}\right)^{1/4} y\,.
\end{align}
Such a possibility is dependent on whether reheating is strong, moderate or weak. 
Note also that $T_{\rm max} \approx 1.2\, T_{\rm reh}\, y^{-1/2}$ and $y \approx 3.2 \,\lambda_6^{10}$ when $\langle \Phi \rangle \ll m_s$. In the next Section we focus on the strong reheating scenario, with moderate and weak reheating being explored in Section~\ref{sec:cohreh}.

\subsubsection{Strong reheating}\label{sec:strongreh}

 As we discussed in Section~\ref{sec:inctr}, if $\Lambda_c\gtrsim 2.4\,\barM$ and both conditions (\ref{c1}) and (\ref{c2}) are satisfied, thermal fluctuations are capable of becoming larger than the height of the potential barrier, and the flaton loses its coherence. If $T_{\rm max}\gtrsim 0.47\Lambda_c$, or equivalently, if
\begin{flalign}\label{eq:l6strmx}
& \text{($T_{\rm max}\gtrsim 0.47\Lambda_c$)} & \Cen{3}{|\lambda_6^{10}| \gtrsim 1.7\times 10^{-9}\left(\frac{g_{\rm max}}{1545/4}\right)^{1/2}\left(\frac{\Lambda_c}{10 \barM}\right)^2\,,}      &&  
\end{flalign}
reheating proceeds through all of the $F_1 {\bar H}$ final-state channels until the temperature drops below the limit imposed by the magnitude of $\delta V_{\rm eff}$.  If in addition the reheating temperature is larger than $0.47\,\Lambda_c$, then the phase transition is not completed until after the end of reheating, and the entire duration of the decay of the inflaton $s$ occurs in the symmetric phase. In terms of the Yukawa coupling, this condition is equivalent to
\begin{flalign}\label{eq:yminstr}
& \text{($T_{\rm reh} > 0.47 \Lambda_c$)} & \Cen{3}{
\begin{aligned}
|\lambda_6^{10}|\; &\gtrsim\; 0.15\left(\frac{5}{8\pi^4 g_{\rm reh}}\right)^{-1/4}(m_s M_P)^{-1/2}\Lambda_c\,,\\
&\simeq\; 2.7\times10^{-5} \left(\frac{g_{\rm reh}}{1545/4}\right)^{1/4}\left(\frac{\Lambda_c}{10 \barM}\right)\,.
\end{aligned}}      &&  
\end{flalign}

\begin{figure}[!ht]
\centering
    \includegraphics[width=0.9\textwidth]{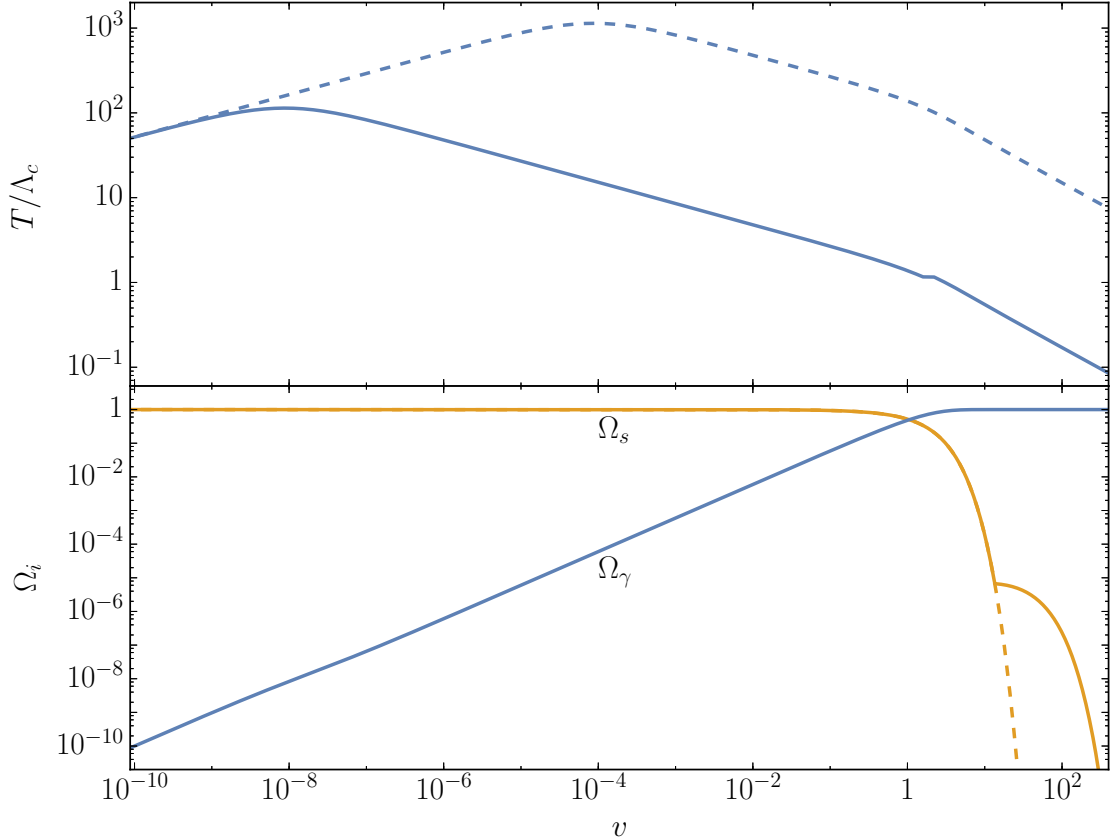}
    \caption{\it Time evolution of the instantaneous temperature and the relative inflaton and radiation densities in the incoherent limit, for $\lambda_6^{10}=10^{-4}$ (solid), $\lambda_6^{10}=10^{-2}$ (dashed) and $\Lambda_c=10\,\barM$.}
    \label{fig:reh_incoh_1}
\end{figure}
 Fig.~\ref{fig:reh_incoh_1} shows the evolution of the instantaneous temperature in units of $\Lambda_c$ (upper panel) and the relative inflaton and radiation energy densities
\beq
\Omega_s = \frac{\rho_s}{\rho_s+\rho_{\gamma}}\quad{\rm and}\quad \Omega_{\gamma}=\frac{\rho_{\gamma}}{\rho_s+\rho_{\gamma}}\,,
\eeq
(lower panel) as functions of the dimensionless time 
\beq
v\equiv \Gamma_{s\rightarrow F_1\bar{H}}(t-t_{\rm end})\,, 
\eeq
for $\lambda_6^{10}=10^{-4}$ and $\Lambda_c=10\,\barM$. At early times, $v\sim 10^{-8}$, the instantaneous temperature of the relativistic plasma reaches its maximum temperature, which is well above the phase transition temperature, $T_{\rm max}\simeq 10^2\Lambda_c$. As the decay of the inflaton $s$ into $F_1 \bar{H}$ proceeds, the temperature decreases following the relation (\ref{eq:Tlatereh}), until the energy density $\rho_s$ is mostly depleted, and the universe becomes dominated by radiation at $v\sim 1$. At a slightly later time, $v\simeq 1.5$, the location of the metastable minimum becomes equal in magnitude to the temperature, $\langle\Phi\rangle\simeq T$, and we therefore assume that the number of degrees of freedom changes from its value in the unbroken phase to its value in the broken phase, in accordance with (\ref{eq:dofub}). This results in a slight increase in the value of $T$
(or, more accurately, a slight delay in the decrease of $T$). At an even later time, $v\simeq 13$, when $T\simeq 0.47\Lambda_c$, the condition (\ref{c1}) is satisfied and the GUT phase transition occurs, driven by the incoherent growth of $\Phi$. The decay of the inflaton switches then from the $F_1 \bar{H}$ channel to solely $\nu^c_1 ~ \Phi$, with a smaller decay rate.
As a result, the rapid decrease of the relative density $\Omega_s$ momentarily stops, since the remaining inflaton energy density redshifts more slowly than radiation, $\rho_s\sim a^{-3}$ vs. $\rho_{\gamma}\sim a^{-4}$. Nevertheless, the decay of $s$ becomes almost immediately effective again and the universe does not cease to be dominated by radiation, given that the condition $\Gamma_{s\rightarrow \nu^c_1 \Phi}(t-t_{\rm end}) \sim 1$ corresponds to $v\sim 20$. Note that the transition has no noticeable effect on the evolution of $T$.

As we indicated earlier, when $\lambda_6^{10} \gg m_s/2\langle \Phi \rangle \simeq 1.5 \times 10^{-3}$, all ten final state channels
are cut off when the GUT phase transition occurs.  But for $\lambda_6^{10}$ this large, reheating is complete before the transition.
As an example, we also consider in Fig.~\ref{fig:reh_incoh_1}, the case with
$\lambda_6^{10} = 10^{-2}$ shown by the dashed lines.  
In the upper panel, the temperature is higher (recall $T_{\rm max}$ 
scales as $\lambda_6^{1/2}$) and 
the feature due to the change in degrees of 
freedom disappears.  In the lower panel the second round of 
decay is gone as all ten channels shut down together.

When the weaker condition $T_{\rm max}\gtrsim 0.47\,\Lambda_c \gtrsim T_{\rm reh}$ is realized, the SU(5)$\times$U(1) phase transition takes place during reheating, and the decay channel (\ref{Glate}) switches on, while the decay (\ref{eq:decvan}) switches off. For strong reheating, in terms of the Yukawa coupling, this regime takes place if
\begin{flalign} \notag
& (T_{\rm max} \gtrsim 0.47\,\Lambda_c \gtrsim T_{\rm reh}) &\\  \label{eq:l6cons}
& \hspace{20pt} 2.7\times10^{-5} \left(\frac{g_{\rm reh}}{1545/4}\right)^{1/4}\left(\frac{\Lambda_c}{10 \barM}\right) \;\gtrsim\; |\lambda_6^{10}| \;\gtrsim\; 1.7\times 10^{-9} \left(\frac{g_{\rm max}}{1545/4}\right)^{1/2} \left(\frac{\Lambda_c}{10\barM}\right)^2\,. &
\end{flalign}
The flaton will be sequestered in the symmetric minimum until the temperature falls below $\Lambda_c$. In the absence of coherence, the vev of the flaton will suddenly transition to the symmetry-breaking minimum as soon as $T\lesssim 0.47\, \Lambda_c$.

\begin{figure}[!t]
\centering
    \includegraphics[width=0.9\textwidth]{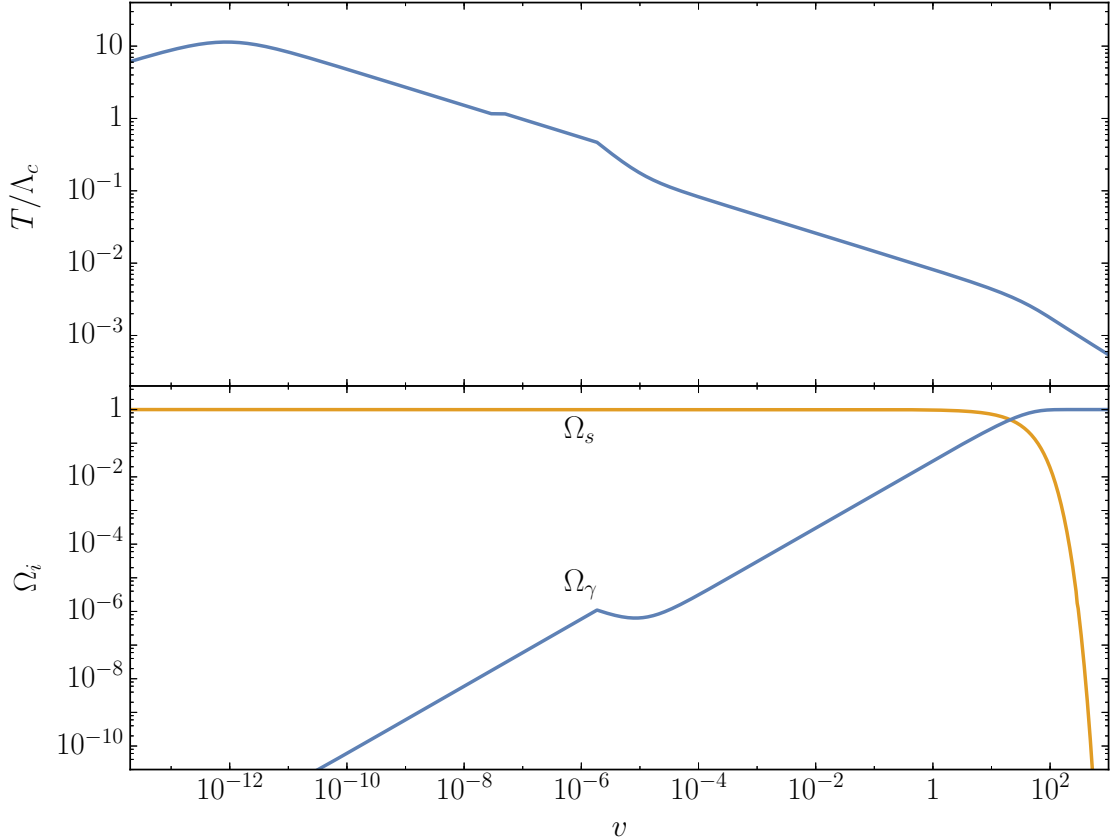}
    \caption{\it Time evolution of the instantaneous temperature and the relative inflaton and radiation densities in the incoherent limit, for $\lambda_6^{10}=10^{-6}$ and $\Lambda_c=10\,\barM$.}
    \label{fig:reh_incoh_2}
\end{figure}
Fig.~\ref{fig:reh_incoh_2} depicts the time-dependence of the temperature and the energy densities of the inflaton and radiation for $\lambda_6^{10}=10^{-6}$ and $\Lambda_c=10\barM$, well within the interval given by (\ref{eq:l6cons}). As expected, the maximum temperature is above the transition threshold, $T_{\rm max}\simeq 11\,\Lambda_c$. In the absence of a phase transition, the decay of the inflaton would be approximately complete around $v\sim1$, as can be checked by extrapolating the trend shown in the lower panel for $\Omega_{\gamma}$ before the GUT $\rightarrow$ SM transition occurs (i.e., $\Omega_\gamma$ would have risen to $\sim 1$ at $v \sim 1$). When this transition happens (at $v \sim 10^{-6}$), the flaton is driven incoherently to the SM minimum, and the decay of $s$ switches from $F_1 \bar{H}$ to $\nu^c_1 \Phi$ production, with a smaller rate. As a result, the previously created radiation density is diluted away by expansion until the inflaton decay ``catches up'' at $v\sim 10^{-5}$. The upper panel shows the temperature decreasing as $T\sim v^{-2/3}$ in this regime, characteristic of the absence of entropy production in a matter-dominated universe. For $v\gtrsim 10^{-5}$, the decay of the inflaton is efficient again, and reheating is finally completed at $v\sim 20$, resulting in a reheating temperature $T_{\rm reh}\ll \Lambda_c$.\\

If $T_{\rm max}\lesssim 0.47\,\Lambda_c$, the flaton will never be trapped in the symmetric vacuum. If $T_{\rm max}$ satisfies the constraint (\ref{eq:Tstrabove}), which is equivalent to
\begin{flalign} \notag
& (0.47 \Lambda_c > T_{\rm max} > \barM) &\\  \label{eq:l6cons2}
& \hspace{50pt} 1.7\times 10^{-9} \left(\frac{g_{\rm max}}{1545/4}\right)^{1/2} \left(\frac{\Lambda_c}{10\barM}\right)^2 \;\gtrsim\; |\lambda_6^{10}| \;\gtrsim\; 7.8\times 10^{-11} \left(\frac{g_{\rm max}}{1545/4}\right)^{1/2} \,, 
\end{flalign}
the phase transition will occur as soon as the radiation background and the flaton thermalize, assuming that this occurs for $T\sim T_{\rm max}$. However, as we discussed in section~\ref{sec:inctemp}, the instantaneous thermalization of the flaton is unlikely, and in the absence of a non-thermal correction of the effective potential we can only ensure that the breakdown of symmetry will occur incoherently if the condition (\ref{eq:Tstrabove}) is satisfied when the decay products of the inflaton thermalize at the temperature $T_{\rm th}$, which can be $T_{\rm th}\ll T_{\rm max}$~\cite{Harigaya:2013vwa,my,ag}.
We note, however, that a small value of $\lambda_6^{10}$ in this case is disfavored
by the consideration of neutrino masses as discussed in Section~\ref{sec:neutrinos},
and thus the uncertainty coming from a lack of a non-thermal correction of 
the effective potential is practically unproblematic in our scenario.

\begin{figure}[!t]
\centering
    \includegraphics[width=0.9\textwidth]{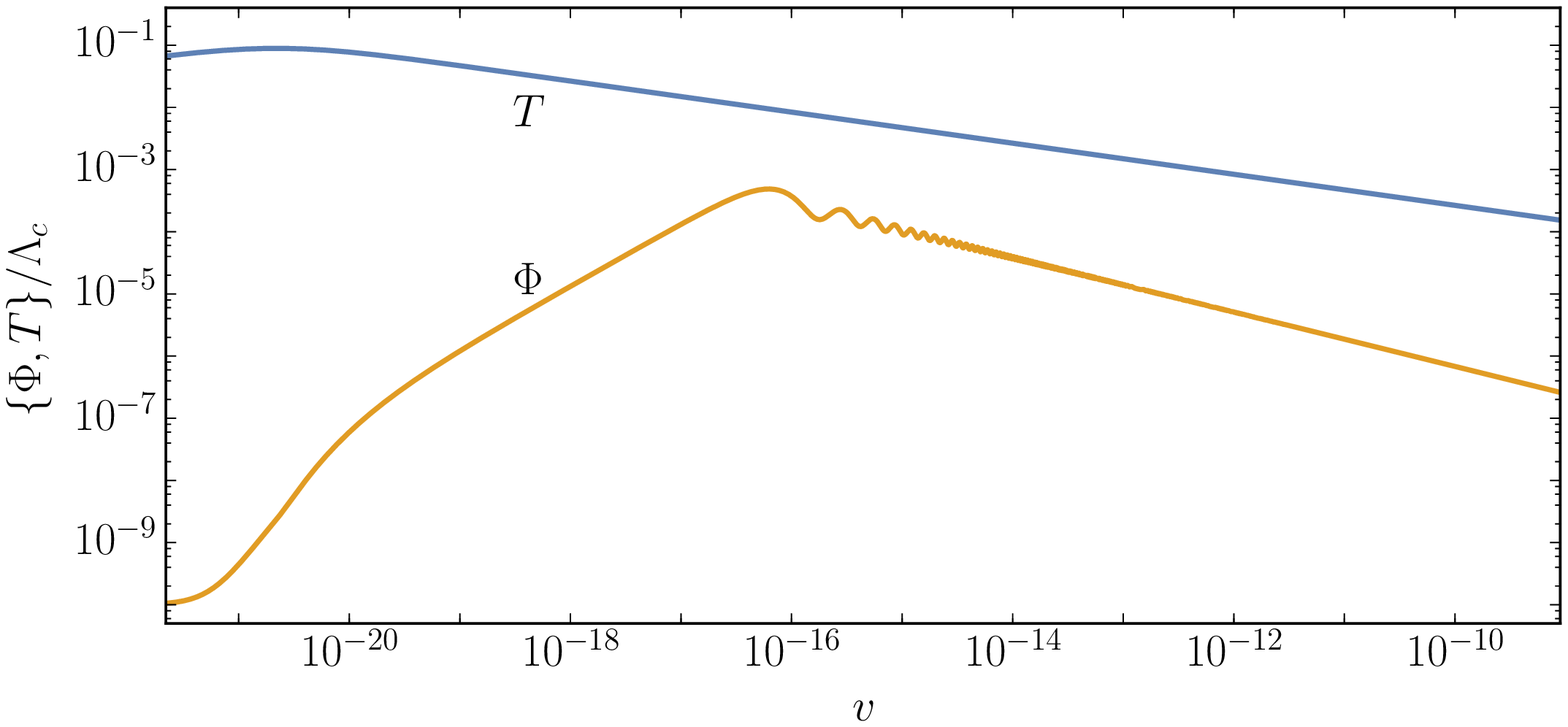}
    \caption{\it Time evolution of the temperature and the flaton in the classical coherent limit for $\lambda_6^{10}=5\times10^{-11}$ and $\Lambda_c=10\,\barM$. }
    \label{fig:reh_stroll}
\end{figure}

For Yukawa couplings $|\lambda_6^{10}| \;\lesssim\; 5.6\times 10^{-11}$, the strong reheating conditions are violated and one must turn to a framework of coherent evolution of the flaton to determine the nature of the transition. As $T_{\rm max}<\Lambda_c$, this case corresponds to the weak reheating scenario~\footnote{Recall that although (\ref{c2}) is satisfied, this is a necessary but not sufficient condition for strong reheating.}.
Fig.~\ref{fig:reh_stroll} shows the time-dependence of $T$ and $\Phi$ for $\lambda_6^{10}=5\times10^{-11}$ and $\Lambda_c=10\,\barM$. The dynamics of $\Phi$ is determined by the solution to the classical equation of motion (\ref{eq:flatonclass}). Starting from $v\ll v_{\rm max}$, the flaton tracks the instantaneous minimum of the effective potential and starts growing. However, at $v\sim 10^{-16}$, this growth is interrupted and the flaton turns around, remaining in the false vacuum because of the presence of a barrier (see Fig.~\ref{fig:veff}). Due to the large curvature of the potential near the barrier, the flaton performs high-frequency oscillations that make it difficult to track numerically its evolution until the end of reheating and beyond. It is therefore unclear if the flaton remains trapped in this false vacuum, or if its appearance is merely a manifestation of the number of approximations made in constructing the effective potential $V_{\rm eff}$. At any rate, at face value, it appears that the decay of the inflaton remains controlled by the $F_1 \bar{H}$ channel even for $T\ll \Lambda_c$, and the $\nu^c_1 \Phi$ channel will dominate only during the later stages of reheating, if at all. If the flaton is indeed trapped, symmetry breaking must occur through tunneling and (in all likelihood) excludes this as a viable scenario.
We again note that this case is disfavored by neutrino masses due to a small $\lambda_6^{10}$ and thus the uncertainty in the flaton evolution is practically insignificant.\\

As we have discussed in this Section, if the strong reheating constraints (\ref{c1}) and (\ref{c2}) are satisfied, the energy density of the flaton is dominated by incoherent thermal fluctuations. After the phase transition is completed, the temperature of the universe continues to decrease, and the interactions that kept the flaton in thermal equilibrium cease to be efficient. If we denote the decoupling temperature as $T_{\rm dec}$, the non-equilibrium temperature profile of $\Phi$ evolves as $T_{\Phi} = T_{\rm dec} (a_{\rm dec}/a) = T (g(T)/g_{\rm dec})^{1/3}$, with $T$ the temperature of the radiation background. At even later times, when $m_{\Phi}>T_{\Phi}$, the flaton becomes non-relativistic, and the universe eventually becomes matter-dominated until the decay of $\Phi$. The Hubble parameter during $\Phi$ domination is then given by
\beq
H \;=\; \left(\frac{\rho_{\Phi}}{3M_P^2}\right)^{1/2} \;=\; \left(\frac{\zeta(3)m_{\Phi}T_{\Phi}^3}{3\pi^2 M_P^2}\right)^{1/2}\,.
\eeq
The decay rate of the flaton was calculated in~\cite{Campbell:1987eb}. It proceeds via effective $D$-term diagrams, leading to the decay rate
\beq
\Gamma_{\Phi}\;\simeq\; \frac{9\lambda_{1,2,3,7}^4}{2048\pi^5}\left(\frac{m_{\Phi}m_{F,\bar{f},\ell^c,\tilde{\phi}_a}^2}{M_{\rm GUT}^2}\right)\,.
\eeq
The flaton decays approximately when $H\sim\Gamma_{\Phi}$, or equivalently when the flaton temperature is
\beq
T_{d\Phi} \;\simeq\; \frac{3\lambda_{1,2,3,7}^{8/3}}{128}\left(\frac{9}{2\zeta(3)\pi^8}\right)^{1/3}\left(\frac{m_{\Phi}m_{F,\bar{f},\ell^c,\tilde{\phi}_a}^4 M_P^2}{M_{\rm GUT}^4}\right)^{1/3}\,.
\eeq
The temperature of the relativistic decay products of $\Phi$ will in turn be given by
\beq
T_{\rm reh}'\;\simeq\; \left(\frac{40}{\pi^2 g_{d\Phi}}\right)^{1/4}(\Gamma_{\Phi}M_P)^{1/2}\;=\; \frac{3 \lambda_{1,2,3,7}^2}{16\pi^3} \left(\frac{5}{8 g_{d\Phi}}\right)^{1/4} \left(\frac{m_{\Phi}m_{F,\bar{f},\ell^c,\tilde{\phi}_a}^2 M_P}{M_{\rm GUT}^2}\right)^{1/2} \,.
\eeq
Here we have neglected the delay arising from the conversion of the heavy supersymmetric decay products into the truly relativistic Standard Model particles. Since the ratio $T_{\rm reh}'/T_{d\Phi}>1$ for $m_{F,\bar{f},\ell^c,\tilde{\phi}_a}\gtrsim 10\,{\rm TeV}$ and $\lambda_{1,2,3,7}\lesssim 1$, the decay of the incoherent flaton results in a net increase of the entropy of the radiation background. The amount of entropy released can be estimated as follows,
\begin{align}\notag
\Delta \;&\equiv\; \frac{s_{\Phi}}{s_{\gamma}}\Bigg|_{d\Phi} \;\simeq\; \frac{g_{d\Phi}T_{\rm reh}^{\prime 3}}{g_{\rm dec}T_{d\Phi}^3}\\ 
&\simeq\; 1.6\times 10^{4}\,\lambda_{1,2,3,7}^{-2}\left(\frac{g_{d\Phi}}{43/4}\right)^{1/4} \left(\frac{915/4}{g_{\rm dec}}\right)\left(\frac{M_{\rm GUT}}{10^{16}\,{\rm GeV}}\right)\left(\frac{10\,{\rm TeV}}{m_{F,\bar{f},\ell^c,\tilde{\phi}_a}^2/m_{\Phi}}\right)^{1/2}\,.
\label{strongD}
\end{align}
This large amount of dilution will have important consequences for the present-day baryon asymmetry, the gravitino decay products and CMB observables, as we discuss further below. Notice that $\Delta$ is independent of the inflaton decay rate, more specifically of $\lambda_6^{10}$. Note also that for $m_{\Phi},m_{F,\bar{f},\ell^c,\tilde{\phi}_a}\gtrsim 10\,{\rm TeV}$, $T_{\rm reh}'\gtrsim 1\,{\rm MeV}$, around what is needed to re-start nucleosynthesis.

\subsubsection{Moderate (weak) reheating}\label{sec:cohreh}

If $\Lambda_c\lesssim 2.4\,\barM$, thermal fluctuations are incapable of driving the breakdown of the $SU(5)\times U(1)$ symmetry, and in this subsection we take $\Lambda_c = 0.4 \barM$ as an example. In this case, the transition is completed by the classical rollover of the coherent flaton $\Phi$ down to its global minimum. As we discussed in Section~\ref{sec:cohdef}, in this case the GUT phase transition
is of the second order, if completed. For $T_{\rm max}\gtrsim \Lambda_c$, the transition is guaranteed to be completed during or after reheating (moderate reheating), while for $T_{\rm max}\lesssim \Lambda_c$, the flaton may remain trapped in the false vacuum until well after reheating is complete (weak reheating).

In analogy with the strong reheating scenario, if $T_{\rm reh}\gtrsim 0.03\Lambda_c$, or equivalently, if
\begin{flalign}\label{eq:corehl1}
& \text{($T_{\rm reh}\gtrsim 0.03\Lambda_c$)} & \Cen{3}{|\lambda_6^{10}| \;\gtrsim \; 7\times 10^{-8} \left(\frac{g_{\rm reh}}{1545/4}\right)^{1/4}\left(\frac{\Lambda_c}{0.4 \barM}\right)\,,}      &&  
\end{flalign}
the decay of the inflaton is completed before the phase transition occurs; at all times $v<1$ the decay proceeds through all of the $F_1 \bar{H}$ final-state channels. Due to the finite size of the barrier $\delta V_{\rm eff}$, the rollover of $\Phi$ to its true low-energy vacuum is delayed until well after $V_{\rm eff}^{\prime\prime}/H^2 \sim 1$. More precisely, in the radiation-dominated universe, the phase transition will be completed when $T\simeq T_{\rm reh}v^{-1/2}\simeq T_{\rm reh}(a_{\rm reh}/a) \sim 0.03\Lambda_c$. With the Hubble parameter during radiation domination given by
\beq
H \;\simeq\; \frac{\Gamma_s}{2}\left(\frac{a_{\rm reh}}{a}\right)^2\,,
\eeq
and denoting by the subindex ``$\Phi$'' the quantities evaluated at the transition time, we obtain the following result for the Hubble-to-mass ratio,
\beq
\frac{H_{\Phi}}{m_{\Phi}} \;\simeq\; \frac{1}{2}\left(\frac{\Gamma_s}{m_{\Phi}}\right)\left(\frac{a_{\rm reh}}{a_{\Phi}}\right)^2 \;\simeq\; \frac{1}{2}\left(\frac{\Gamma_s}{m_{\Phi}}\right)\left(\frac{0.03\Lambda_c}{T_{\rm reh}}\right)^2 \;\simeq\; 2\times 10^{-6} \left(\frac{10\,{\rm TeV}}{m_{\Phi}}\right)\left(\frac{\Lambda_c}{0.4\,\barM}\right)^{2}\,.
\eeq
This ratio is much smaller than one {\em independently of the value of $|\lambda_6^{10}|$}. This implies that, well before the phase transition, the flaton does not adiabatically track the position of the instantaneous minimum, but instead tracks it while continuously oscillating about it. This in particular makes it numerically challenging to follow the dynamics beyond $v\sim1$. Moreover, following the transition, the flaton begins large amplitude oscillations about its minimum, which are very underdamped, and which begin much later than the naive estimate $H_{\Phi}\sim m_{\Phi}$. Therefore, it is to be expected that these oscillations will eventually dominate the energy budget of the universe until the decay of the flaton, the moment at which a large amount of entropy will be released. We can estimate the amount of entropy released by noting that, at $\Phi$-radiation equality, the flaton and radiation energy densities will be given by
\beq
\begin{aligned}
\rho_{\Phi}\;&\simeq\; V_0\left(\frac{a_*}{a_\Phi}\right)^{-3}\,,\\
\rho_{\gamma} \;&\simeq\; \frac{3}{4}\Gamma_s^2M_P^2\left(\frac{a_*}{a_{\rm reh}}\right)^{-4}\,,
\end{aligned}
\eeq
where the scale factor at equality is denoted by $a_*$. We can then evaluate the ratio
\beq
\frac{a_*}{a_{\Phi}} \;\sim\; \left(\frac{\Gamma_s M_P}{m_{\Phi}M_{\rm GUT}}\right)^2\left(\frac{a_{\rm reh}}{a_{\Phi}}\right)^4\; \simeq \;  \left(\frac{\Gamma_s M_P}{m_{\Phi}M_{\rm GUT}}\right)^2\left(\frac{0.03\Lambda_c}{T_{\rm reh}}\right)^4\,.
\eeq
Note that because of the delay in the phase transition, 
$a_\Phi$ is much larger than it would have been had the transition occurred when $H \sim m_\Phi$. As a result, $a_*/a_{\Phi}$ is greatly reduced and, as we will see, the amount of entropy production will be greatly enhanced. 

The Hubble parameter during $\Phi$ domination is given by
\beq
H\;\sim \; \frac{m_{\Phi}M_{\rm GUT}}{M_P}\left(\frac{a_{\Phi}}{a}\right)^{3/2}\,.
\eeq
Since the decay of the flaton occurs  at $H\sim \Gamma_{\Phi}$, we can compute the energy density ratio at decay as follows,
\beq
\frac{\rho_{\Phi}}{\rho_{\gamma}}\Bigg|_{d\Phi} \;=\; \left(\frac{a_{d\Phi}}{a_*}\right) \;=\; \left(\frac{a_{d\Phi}}{a_{\Phi}}\right)\left(\frac{a_{\Phi}}{a_*}\right) \sim \left(\frac{m_{\Phi}^4M_{\rm GUT}^4}{\Gamma_{\Phi}\Gamma_s^3M_P^4}\right)^{2/3}\left(\frac{T_{\rm reh}}{0.03\Lambda_c}\right)^4\,.
\eeq
With the entropy density in radiation given by $s_{\gamma} = 4/3(g_{\rm reh}\pi^2/30)^{1/4}\rho^{3/4}_{\gamma}$, and a similar
expression for the entropy density produced from $\Phi$ decays, we can finally evaluate the entropy release due to the decay of the flaton
\begin{align} \displaybreak[0]
\Delta \;&\sim\; \left(\frac{g_{d\Phi}}{g_{\rm reh}}\right)^{1/4} \left(\frac{m_{\Phi}^4M_{\rm GUT}^4}{\Gamma_{\Phi}\Gamma_s^3M_P^4}\right)^{1/2}\left(\frac{T_{\rm reh}}{0.03\Lambda_c}\right)^{3} \\ \notag
&\simeq\; 4.8\times 10^{17}\, \lambda_{1,2,3,7}^{-2} \left(\frac{g_{d\Phi}}{g_{\rm reh}}\right)^{1/4}  \left(\frac{m_{\Phi}}{10\,{\rm TeV}}\right)^{3/2}\left(\frac{m_{F,\bar{f},\ell^c,\tilde{\phi}_a}}{10\,{\rm TeV}}\right)^{-1} \\ \label{eq:entropyv1}
&\qquad \times \left(\frac{M_{\rm GUT}}{10^{16}\,{\rm GeV}}\right)^3  \left(\frac{\Lambda_c}{0.4\barM}\right)^{-3}\,.
\end{align}
Note that this estimate is independent of the value of $|\lambda_6^{10}|$, indicating that an enormous amount of dilution due to entropy release is to be expected following the decay of $\Phi$ if the phase transition is completed following the end of reheating. 
Given the enormous dilution factor in (\ref{eq:entropyv1}),
we do not display the evolution of the fields for this case.

If $T_{\rm max}\gtrsim 0.03\Lambda_c\gtrsim T_{\rm reh}$ one would expect the GUT phase transition to be completed coherently during reheating. However, unless $T_{\rm max}\gtrsim \Lambda_c$, the flaton will not have a chance to roll from the false vacuum toward the global minimum before getting trapped by the barrier that appears for $T\ll \Lambda_c$, as we discussed in sections~\ref{sec:cohdef} and \ref{sec:strongreh}. In terms of the Yukawa coupling, this moderate reheating regime is realized for
\begin{flalign} \notag
& (T_{\rm max}\gtrsim \Lambda_c;~~ 0.03\Lambda_c\gtrsim T_{\rm reh}) &\\  \label{eq:l6truemod}
& \hspace{40pt} 7\times 10^{-8} \left(\frac{g_{\rm reh}}{1545/4}\right)^{1/4} \left(\frac{\Lambda_c}{0.4 \barM}\right) \;\gtrsim \;  |\lambda_6^{10}| \;\gtrsim\; 1.2\times 10^{-11} \left(\frac{g_{\rm max}}{1545/4}\right)^{1/2} \left(\frac{\Lambda_c}{0.4 \barM}\right)^2 \,. &
\end{flalign}
When this is the case, the reheating temperature is given by (\ref{eq:Trehstr}) with $y=y_6^\Phi \simeq \lambda_6^{10}/\sqrt{2}$ as per (\ref{y6late}). Therefore, the temperature between $T_{\rm max}$ and $T_{\rm reh}$ will be related to the reheating temperature by
\beq
T \;\simeq \; \begin{cases}
\left(\dfrac{a_{\rm reh}}{a}\right)^{3/8}\left(\dfrac{y_6^0}{y_6^\Phi}\right)^{1/2}T_{\rm reh}\,, & a<a_{\Phi}\,,\\[10pt] 
\left(\dfrac{a_{\rm reh}}{a}\right)^{3/8} T_{\rm reh}\,, & a \gg a_{\Phi}\,.
\end{cases}
\label{cases}
\eeq
The factor $(y_6^0/y_6^\Phi)^{1/2} \simeq (20)^{1/4}$ accounts for the fact that before the transition the inflaton decay rate is proportional to ${y_6^0}^2$ whereas after
the transition which occurs at $a_\Phi$, the decay rate is reduced to 
being proportional to ${y_6^\Phi}^2$ and less energy is pumped into the radiation bath. Recall also that $T_{\rm reh}$ is now defined in terms of $y_6^\Phi$.
Similarly, when $\delta V_{\rm eff}\rightarrow 0$ and $\Phi>m_s$,
\beq
\frac{|V_{\rm eff}^{\prime\prime}|}{H^2}\;\simeq\; \frac{9m_{\Phi}^2 v_{\Phi}^2}{4\Gamma_s^2} \;\simeq\; \frac{9}{4}\left(\frac{a}{a_{\rm reh}}\right)^3\left(\frac{m_{\Phi}M_P}{T_{\rm reh}^2}\right)^2 \left(
\frac{40}{\pi^2 g_{\rm reh}}\right)\,,
\eeq
where we have used $H^2 = (4/9) \Gamma_s^2 (a_{\rm reh}/a)^2$ corresponding to expansion dominated by inflaton oscillations, and where $v_{\Phi}=\Gamma_{s\rightarrow \nu^c_1 \Phi}(t-t_{\rm end})$.
The rollover of the flaton $\Phi$ to its true low-energy vacuum will occur when $T\simeq 0.03\Lambda_c$, or when $|V_{\rm eff}^{\prime\prime}|^{1/2}/H \sim 10$, whichever happens later.\footnote{The condition $|V_{\rm eff}^{\prime\prime}|^{1/2}/H \sim \mathcal{O}(10)$ is used to match our
numerical integrations. We found that $|V_{\rm eff}^{\prime\prime}|^{1/2}/H \sim \mathcal{O}(1)$ in Eq.~(\ref{eq:flatonclass}) instead describes an overdamped oscillator.} 
That is, we use (\ref{cases}) with $a < a_\Phi$. However for small $m_\Phi$, the transition may be further delayed and thus we must compare
both determinations of $a_\Phi$. 
In the light of the previous two equations, this is equivalent to
\beq
\frac{a_{\Phi}}{a_{\rm reh}}\;\simeq\; \max\left[ \left(\frac{y_6^0}{y_6^\Phi}\right)^{4/3} \left(\frac{T_{\rm reh}}{0.03\,\Lambda_c}\right)^{8/3}\,,\, \left(\frac{50T_{\rm reh}^2}{m_{\Phi}M_P}\right)^{2/3}\right]\,,
\eeq
where we have used $g_{\rm reh} = 915/4$. 
Assuming that the oscillations of $\Phi$ eventually dominate over the radiation background until the decay of the flaton, we can then write
\begin{align}\notag
\frac{\rho_{\Phi}}{\rho_{\gamma}}\Bigg|_{d\Phi} \;&=\; \left(\frac{a_{d\Phi}}{a_{\Phi}}\right)\left(\frac{a_{\Phi}}{a_*}\right)\\
& \sim\; \left(\frac{m_{\Phi}^4M_{\rm GUT}^4}{\Gamma_{\Phi}\Gamma_s^3 M_P^4}\right)^{2/3} \max\left[ \left(\frac{y_6^0}{y_6^\Phi}\right)^{16/3} \left(\frac{T_{\rm reh}}{0.03\,\Lambda_c}\right)^{32/3}\,,\,  \left(\frac{50T_{\rm reh}^2}{m_{\Phi}M_P}\right)^{8/3}\right]\,,
\end{align}
to obtain finally  the amount of entropy dilution due to the decay of $\Phi$,
\beq
\Delta \;\sim\; \left(\frac{g_{d\Phi}}{g_{\rm reh}}\right)^{1/4}  \left(\frac{m_{\Phi}^4M_{\rm GUT}^4}{\Gamma_{\Phi}\Gamma_s^3 M_P^4}\right)^{1/2} \max\left[ \left(\frac{y_6^0}{y_6^\Phi}\right)^{4} \left(\frac{T_{\rm reh}}{0.03\,\Lambda_c}\right)^{8}\,,\, \left(\frac{50T_{\rm reh}^2}{m_{\Phi}M_P}\right)^{2}\right]\,.
\eeq
For $3.3\times 10^{-10}\lesssim |\lambda_6^{10}|\lesssim 7\times 10^{-8}$ this results in 
\begin{align} \notag
\Delta \;&\sim\; 2.1\times 10^{17}\, \lambda_{1,2,3,7}^{-2} \left(\frac{g_{d\Phi}}{g_{\rm reh}}\right)^{1/4} \left(\frac{|\lambda_6^{10}|}{7\times 10^{-8}}\right)^5 \left(\frac{m_{\Phi}}{10\,{\rm TeV}}\right)^{3/2}\left(\frac{m_{F,\bar{f},\ell^c,\tilde{\phi}_a}}{10\,{\rm TeV}}\right)^{-1} \\ \label{eq:entropyv2}
&\qquad \times \left(\frac{M_{\rm GUT}}{10^{16}\,{\rm GeV}}\right)^3 \left(\frac{m_s}{3\times 10^{13}\,{\rm GeV}}\right)^{5/2} \left(\frac{\Lambda_c}{4\times 10^9\,{\rm GeV}}\right)^{-8}\,,
\end{align}
while for $1.2\times 10^{-11}\lesssim |\lambda_6^{10}|\lesssim 3.3\times 10^{-10}$ we obtain
\begin{align} \notag
\Delta \;&\sim\; 1.1\times 10^{8}\, \lambda_{1,2,3,7}^{-2} \left(\frac{g_{d\Phi}}{g_{\rm reh}}\right)^{1/4} \left(\frac{|\lambda_6^{10}|}{7\times 10^{-8}}\right) \left(\frac{m_{\Phi}}{10\,{\rm TeV}}\right)^{-1/2}\left(\frac{m_{F,\bar{f},\ell^c,\tilde{\phi}_a}}{10\,{\rm TeV}}\right)^{-1} \\ \label{eq:entropyv3}
&\qquad \times \left(\frac{M_{\rm GUT}}{10^{16}\,{\rm GeV}}\right)^3 \left(\frac{m_s}{3\times 10^{13}\,{\rm GeV}}\right)^{1/2}\,.
\end{align}\par

\begin{figure}[!t]
\centering
    \includegraphics[width=0.9\textwidth]{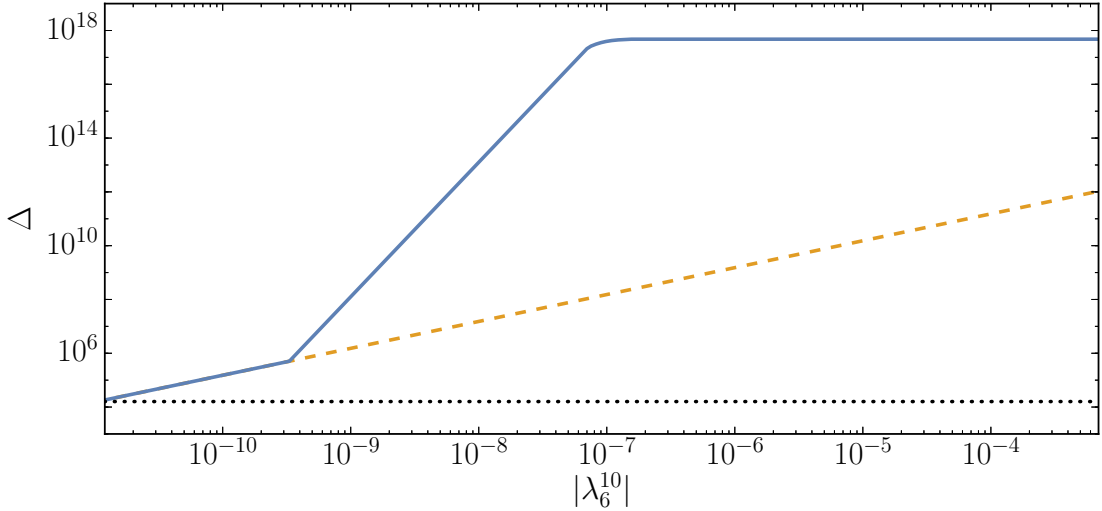}
    \caption{\it Entropy dilution factor for moderate reheating as a function of $\lambda_6^{10}$. Continuous line: the dilution factor accounting for a delayed phase transition, Eqs.~(\ref{eq:entropyv1}), (\ref{eq:entropyv2}) and (\ref{eq:entropyv3}). Dashed line: the dilution factor ignoring the delayed phase transition, Eq.~(\ref{eq:entropyv3}). The horizontal dotted line corresponds to the entropy production
    in the strong reheating case with $\Lambda_c = 10\barM$ from Eq. (\ref{strongD}).}
    \label{fig:entropy}
\end{figure}

Fig.~\ref{fig:entropy} shows the dependence of the entropy dilution factor $\Delta$ on the Yukawa coupling $\lambda_6^{10}$ for moderate reheating. The solid curve corresponds to the full result given by Eqs.~(\ref{eq:entropyv1}), (\ref{eq:entropyv2}) and (\ref{eq:entropyv3}), while the dashed curve corresponds only to (\ref{eq:entropyv3}), which ignores the delay in the phase transition due to the presence of the barrier of height $\delta V_{\rm eff}$ for $T\lesssim 0.03\Lambda_c$. This latter result corresponds to that calculated in~\cite{egnno2}. Note that, as expected, the amount of entropy production is smallest in the case when the transition happens close to $v_{\rm max}$. For $|\lambda_6^{10}|\gtrsim 3\times 10^{-10}$, the dilution factor is enhanced due to the delay of the transition, as the flaton starts oscillations late, resulting in a large energy density for $\Phi$ at the moment of its decay. Recall that for 
$|\lambda_6^{10}| < 1.2\times 10^{-11}$, the transition is never completed.

\par

Also shown in Fig.~\ref{fig:entropy} is the entropy production in the
strong reheating case taken from Eq.~(\ref{strongD}) with $\Lambda_c = 10 \barM$. As one can see, the dilution factor is always larger in the moderate reheating case (with $\Lambda_c = 0.4 \barM$) than that in the strong reheating scenario, inevitably leading to the severe washout out of any prior baryon 
asymmetry. Note also that the strong reheating result (\ref{strongD}) is independent of $\Lambda_c$ and therefore serves as an estimate of the entropy released by the decay of the thermal flatons produced by inflaton decay. \\

\begin{figure}[!t]
\centering
    \includegraphics[width=0.9\textwidth]{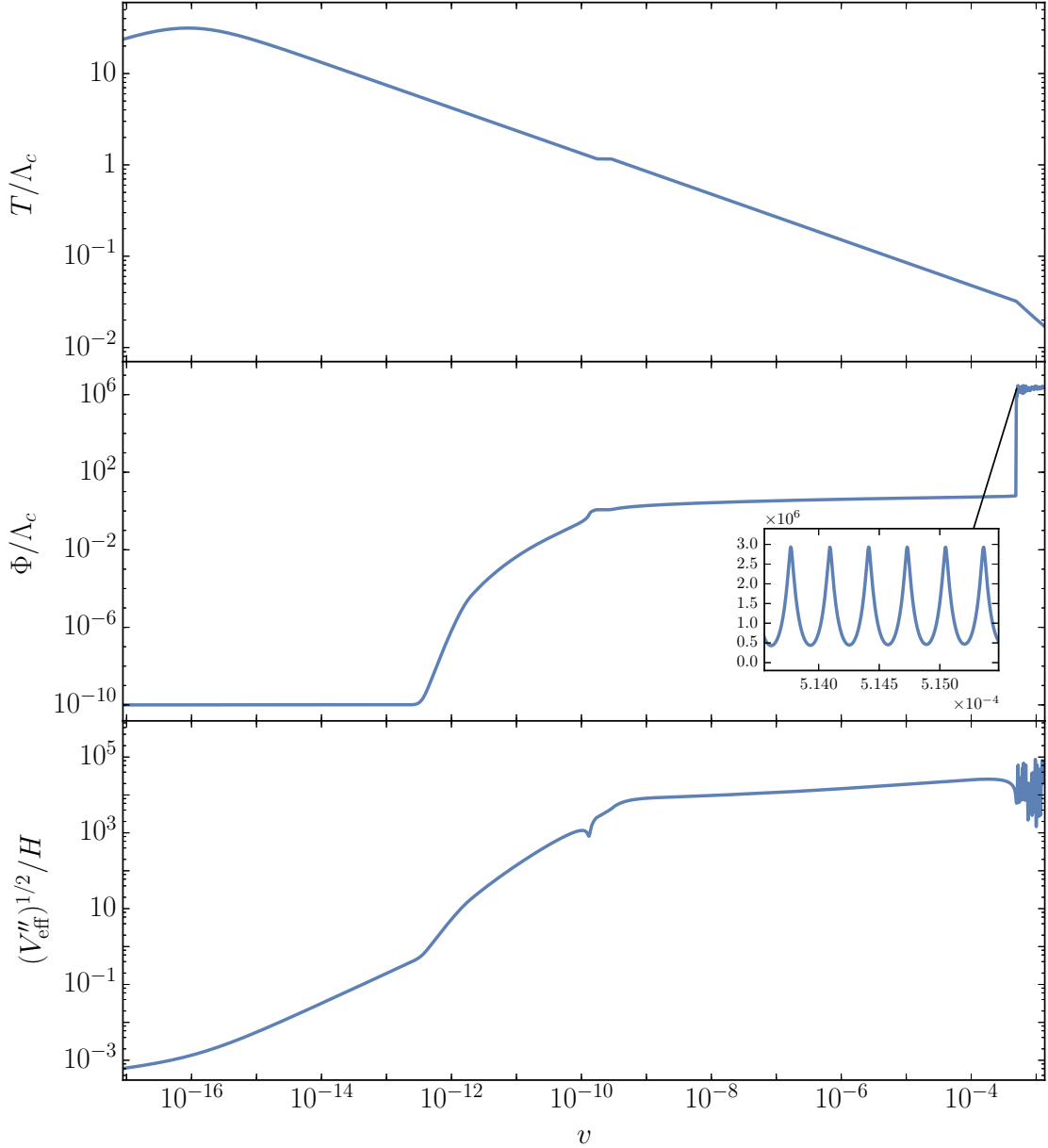}
    \caption{\it Time evolution of the temperature, the flaton and the ratio of the instantaneous curvature of the potential to the Hubble parameter, in the classical coherent limit with $\lambda_6^{10}=10^{-8}$ and $\Lambda_c=0.4\,\barM$. The inset shows the detail of the oscillations of the $\Phi$ field about the SM minimum.}
    \label{fig:reh_roll_1}
\end{figure}\par
Fig.~\ref{fig:reh_roll_1} shows the classical evolution of the temperature, the flaton $\Phi$ and the Hubble parameter for $\lambda_6^{10}=10^{-8}$ and $\Lambda_c=0.4\,\barM$, illustrating the regime for which (\ref{eq:entropyv2}) applies. At very early times ($v \sim 10^{-12}$), the flaton moves from its assumed starting point of $\Phi/\Lambda_c = 10^{-10}$ to $\Phi/\Lambda_c \sim 1$ by tracking its instantaneous local minimum. At $v\simeq 10^{-10}$, $T=\Phi$ and the number of degrees of freedom changes in accordance with (\ref{eq:dofub}). The rate of decrease in $T$ is then slowed down until it agrees with the corresponding equilibrium value. At a later time, $v\simeq 5\times 10^{-4}$, the rapid rollover of the flaton to the SM vacuum is observed; it occurs around $T\sim 0.03\,\Lambda_c$, in agreement with Fig.~\ref{fig:vev}. Note that the sudden rollover is followed by underdamped oscillations of $\Phi$ about the global minimum with large amplitude, as seen in the inset plot. This is explained by the fact that, by the time the transition is completed, the Hubble parameter has decreased by a factor of $\sim10^{13}$ since $T=T_{\rm max}$, and is much smaller than the instantaneous curvature (mass) of the potential. After the transition the temperature decreases with a larger rate, as the decay of $s$,  which is not yet complete, switches from $F_1 \bar{H}$ to the $\nu^c_1 \Phi$ channel. With a smaller decay rate, the radiation energy density present in the Universe is diluted by expansion without significant production, until the decay becomes ``operative'' at later times, similarly to what occurs in the strong reheating case shown in Fig.~\ref{fig:reh_incoh_2}. However, unlike the latter case, the large-amplitude, large-frequency oscillations of the coherent flaton do not allow us to easily track the dynamics of the inflaton-radiation-flaton system up to and beyond the end of reheating.\\

\begin{figure}[!t]
\centering
    \includegraphics[width=0.9\textwidth]{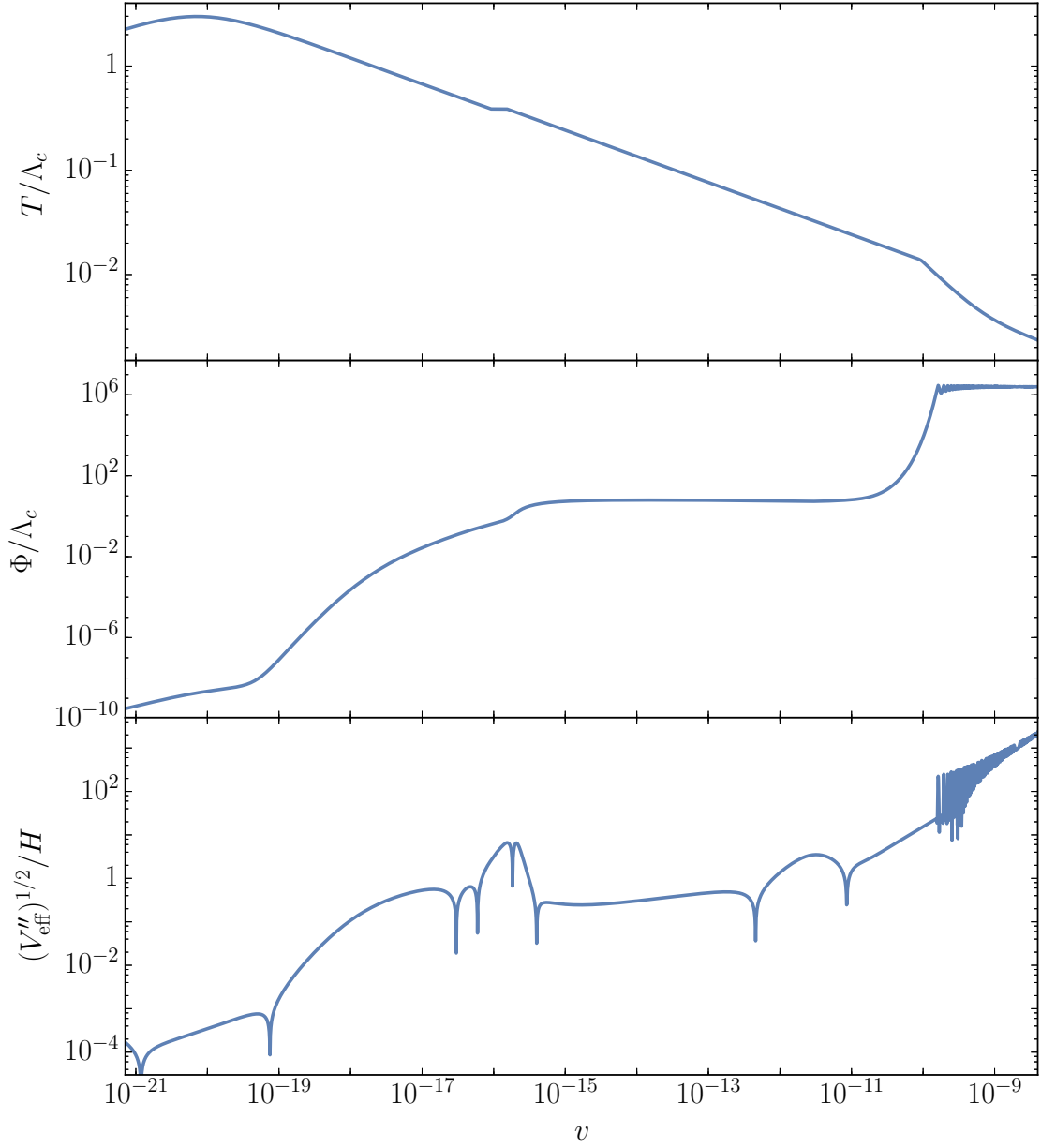}
    \caption{\it Time evolution of the temperature, the flaton and the ratio of the instantaneous curvature of the potential to the Hubble parameter, in the classical coherent limit for $\lambda_6^{10}=9\times10^{-11}$ and $\Lambda_c=0.4\,\barM$. }
    \label{fig:reh_roll_2}
\end{figure}
In Fig.~\ref{fig:reh_roll_2}, the dynamics of the temperature, the coherent flaton and the Hubble parameter are shown for a coupling near the moderate-weak reheating boundary, $\lambda_6^{10}=9\times10^{-11}$, $\Lambda_c=0.4\,\barM$. This corresponds to the regime for which (\ref{eq:entropyv3}) is valid. Note that in this case the rollover of $\Phi$ toward the SM vacuum is gentle, a reflection of the fact that $V_{\rm eff}^{\prime\prime}/H_{\Phi}^2 \sim 1$ during the phase transition. Tracking the oscillating field well after the transition is also challenging for the chosen model parameter values, forcing us to stop the integration well before the end of reheating.\\

\begin{figure}[!ht]
\centering
    \includegraphics[width=0.9\textwidth]{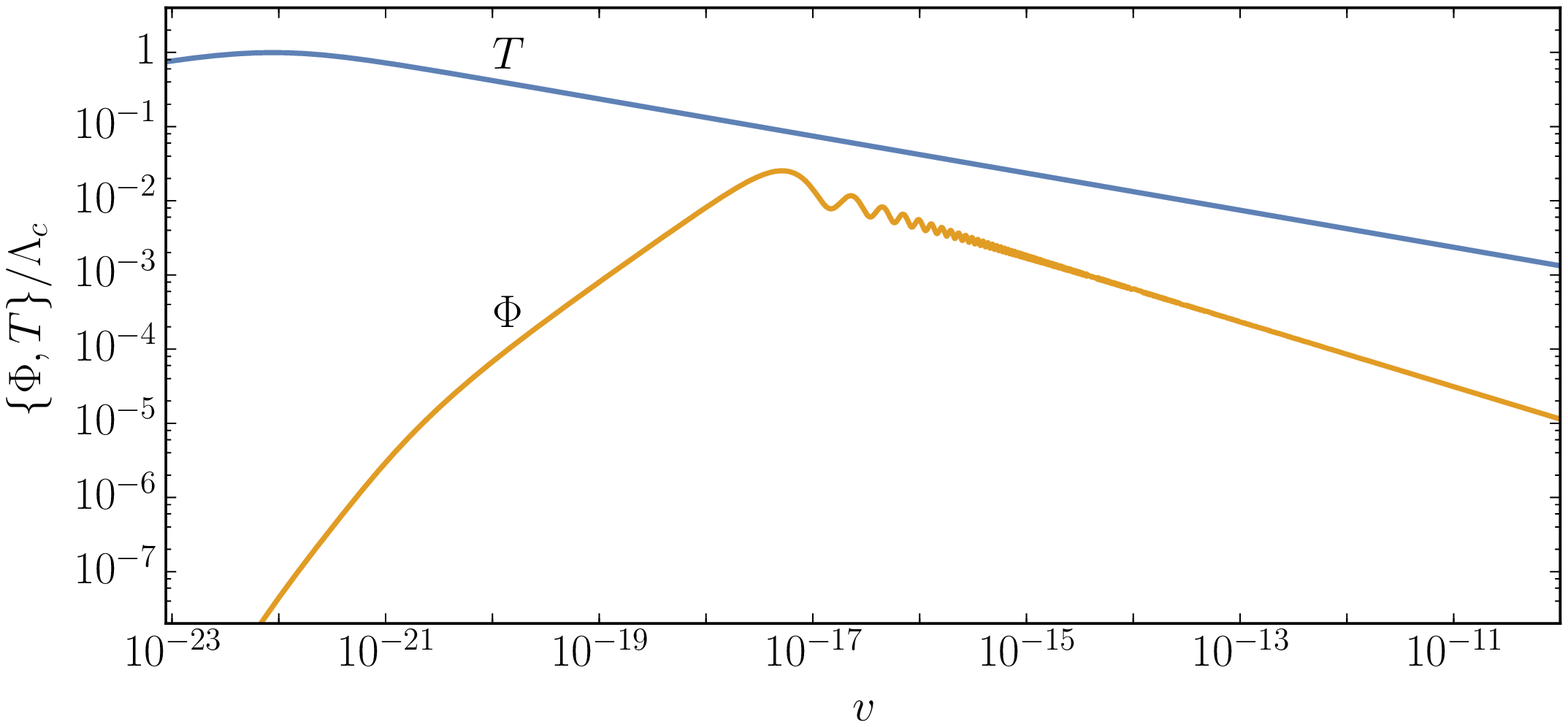}
    \caption{\it Time evolution of the temperature and the flaton in the classical coherent limit for $\lambda_6^{10}=10^{-11}$ and $\Lambda_c=0.4\,\barM$. }
    \label{fig:reh_roll_3}
\end{figure}
For $|\lambda_6^{10}|\lesssim 1.2\times 10^{-11}$, we expect the flaton to remain trapped in the false vacuum even for $T\ll 0.03\Lambda_c$. Fig.~\ref{fig:reh_roll_3} shows the time dependence of $T$ and $\Phi$ for $\lambda_6^{10}=10^{-11}$ and $\Lambda_c=0.4\,\barM$. Note the similar evolution to that shown in Fig.~\ref{fig:reh_stroll}: the flaton tracks the instantaneous minimum of the effective potential but does not grow past the temperature, turning around instead and oscillating in the false vacuum, excluding this as a viable scenario.

\section{Gravitino production}
\label{sec:gravitino}

We now proceed to track the production of gravitinos during reheating \cite{weinberg,elinn,nos,ehnos,kl,ekn,eln,Juszkiewicz:gg,mmy,Kawasaki:1994af,Moroi:1995fs,enor,Giudice:1999am,bbb,kmy,stef,Pradler:2006qh,ps2,rs,kkmy}. Disregarding the finite lifetime of the gravitino for now, the Boltzmann equation for the gravitino number density can be written as
\beq\label{eq:n32eq}
\frac{d n_{3/2}}{dt} + 3Hn_{3/2} \;=\; \langle \sigma v\rangle n_{\rm rad}^2\,,
\eeq
where $n_{\rm rad}=\zeta(3) T^3/\pi^2$ is the number density of any single bosonic relativistic degree of freedom, and where $\langle \sigma v\rangle$ is the thermally-averaged gravitino production cross section. This average cross section can be written as~\cite{bbb,Pradler:2006qh,rs,EGNOP}
\beq
\langle \sigma v\rangle \;=\; \langle \sigma v\rangle_{\rm top} + \langle \sigma v\rangle_{\rm gauge} \,,
\eeq
with
\beq
\langle \sigma v\rangle_{\rm top} \;=\; 1.29 \frac{|y_t|^2}{M_P^2} \left(1+ \frac{A_t^2}{3 m_{3/2}^2}\right)\,,
\eeq
where $A_t$ is the top-quark supersymmetry-breaking trilinear coupling, and
\begin{align}
\langle \sigma v\rangle_{\rm gauge} \;&=\; \sum_{i=1}^3 \frac{3\pi c_i g_i^2}{16\zeta(3) M_P^2} \left(1+\frac{m_{\tilde{g}_i}^2}{3 m_{3/2}^2}\right) \ln\left(\frac{k_i}{g_i}\right)\,,\\ \label{eq:sigmagrav}
&=\; \frac{26.24}{M_P^2}\Bigg[ \left(1+0.558 \frac{m_{1/2}^2}{m_{3/2}^2}\right) - 0.011 \left(1+3.062 \frac{m_{1/2}^2}{m_{3/2}^2}\right) \ln\left(\frac{T}{10^{10}\,{\rm GeV}}\right)\Bigg]\,,
\end{align}
where the $m_{\tilde{g}_i}$ are the gaugino masses and the constants $c_i$, $k_i$ depend on the gauge group (see~\cite{EGNOP} for details). The first term in the gaugino mass-dependent factors $(1+ m_{\tilde{g}_i}^2 / 3 m_{3/2}^2)$ corresponds to the production of the transversely polarized gravitino, while the second term is associated with the production of the longitudinal (Goldstino) component. In what follows we will focus exclusively on the ``gauge'' contribution to the cross section, which dominates. 
It is worth noting that Eq.~(\ref{eq:sigmagrav}) is strictly valid only {\em after} the GUT phase transition. At earlier times, the correct expression depends on the SU(5)$\times$U(1) coupling constants. In the absence of the correct cross section in the unbroken phase, we will make use of (\ref{eq:sigmagrav}) for any $\lambda_6$ for illustrative purposes. Nevertheless, we expect our results to be exact in the case when the phase transition is completed well before the end of reheating.\\

For an inflaton decay rate with {\em constant} Yukawa coupling $y$, the gravitino yield
\beq
Y_{3/2}\;\equiv \; \frac{n_{3/2}}{n_{\rm rad}}\,,
\eeq
can be computed to give, at low temperatures $T\ll 1\,{\rm MeV}$~\cite{EGNOP}
\begin{align}
Y_{3/2}(T)\;&\simeq \;  \Delta^{-1}\,\frac{g(T)}{g_{\rm reh}}\, Y_{3/2}(T_{\rm reh})\\ \label{eq:y32}
&\simeq \; 2.5\times 10^{-6} |y| \Delta^{-1}\left(\frac{915/4}{g_{\rm reh}}\right)\left(1+0.56 \frac{m_{1/2}^2}{m_{3/2}^2}\right)\,,
\end{align}
where the factor $\Delta$ accounts for entropy production from flaton decay. 

Although $R$-parity is violated
in our flipped SU(5)$\times$U(1) model, it is sufficiently sequestered from the observable sector that the
the lifetime of the lightest supersymmetric particle (LSP) is much longer than the age of the universe. For example, if the LSP is a neutralino \cite{ehnos}, the LSP can decay through its Higgsino component $\tilde h$ into $L F F^*$
via the $\lambda_2 F {\bar f} {\bar h}$ coupling, followed by
$\tilde{ \nu^c} - S$ mixing and a $S F F^*$ coupling induced
at 1-loop. We estimate the rate for this decay to be
\begin{equation}
    \Gamma_{\rm LSP} \; \sim \; \frac{1}{192 \pi^3} \,
    \frac{\lambda_2^2 \lambda_6^4 m_s^2}{(16 \pi^2)^2 M_{\rm GUT}^4 M_P^2} \, m_{\rm LSP}^5 \, ,
    \label{GammaLSP}
\end{equation}
corresponding to a lifetime that we estimate to be
in excess of $10^{72}$~years for $\lambda_2 \sim \lambda_6 \sim 10^{-5}$, $\lambda_8 \sim m_s/M_P$ and $m_{\rm LSP} \sim 100$~GeV.
Therefore, the LSP is as good a candidate for cold dark matter as in $R$-conserving models.
The relic density of cold dark matter produced by gravitino decay, assuming $n_{\rm LSP}=n_{3/2}$, can be written as 
\begin{align} \notag
\Omega_{\rm CDM}h^2 \;&=\; \frac{m_{\rm LSP} Y_{3/2}n_{\gamma}}{2\rho_c\, h^{-2}}\\
&\simeq\; 0.12\, \Delta^{-1}\left(\frac{|y|}{2.4\times 10^{-5}}\right)\left(\frac{m_{\rm LSP}}{100\,{\rm GeV}}\right)\left(\frac{915/4}{g_{\rm reh}}\right)\left(1+0.56 \frac{m_{1/2}^2}{m_{3/2}^2}\right)\,,
\label{ocdm}
\end{align}
where $\rho_c=1.054\times 10^{-5} h^2\, {\rm GeV} {\rm cm}^{-3}$ is the closure density, and where the factor of 2 is present because we have defined $Y_{3/2}$ in terms of $n_{\rm rad} = n_{\gamma}/2$. In the absence of entropy production, this leads to the constraint $|y|\lesssim 2.4\times 10^{-5}$ in order to avoid the overabundant production of the LSP. 
An immediate consequence of Eq. (\ref{ocdm}) is that the 
correct relic density can be obtained from gravitino decay
when
\beq
|y| = 2.4 \times 10^{-5} \Delta \left(1+0.56 \frac{m_{1/2}^2}{m_{3/2}^2}\right)^{-1} \left(\frac{100\,{\rm GeV}}{m_{\rm LSP}}\right) \, .
\label{ylimit}
\eeq
For $\Delta \sim 10^{4}$, $y\sim 0.24$ when
the longitudinal modes are not dominant ($m_{1/2} \ll m_{3/2}$). However
$y$ may be much smaller (and in the range considered above) when
either $m_{1/2} \gg m_{3/2}$ or $m_{\rm LSP} > 100$ GeV.

\begin{figure}[!t]
\centering
    \includegraphics[width=\textwidth]{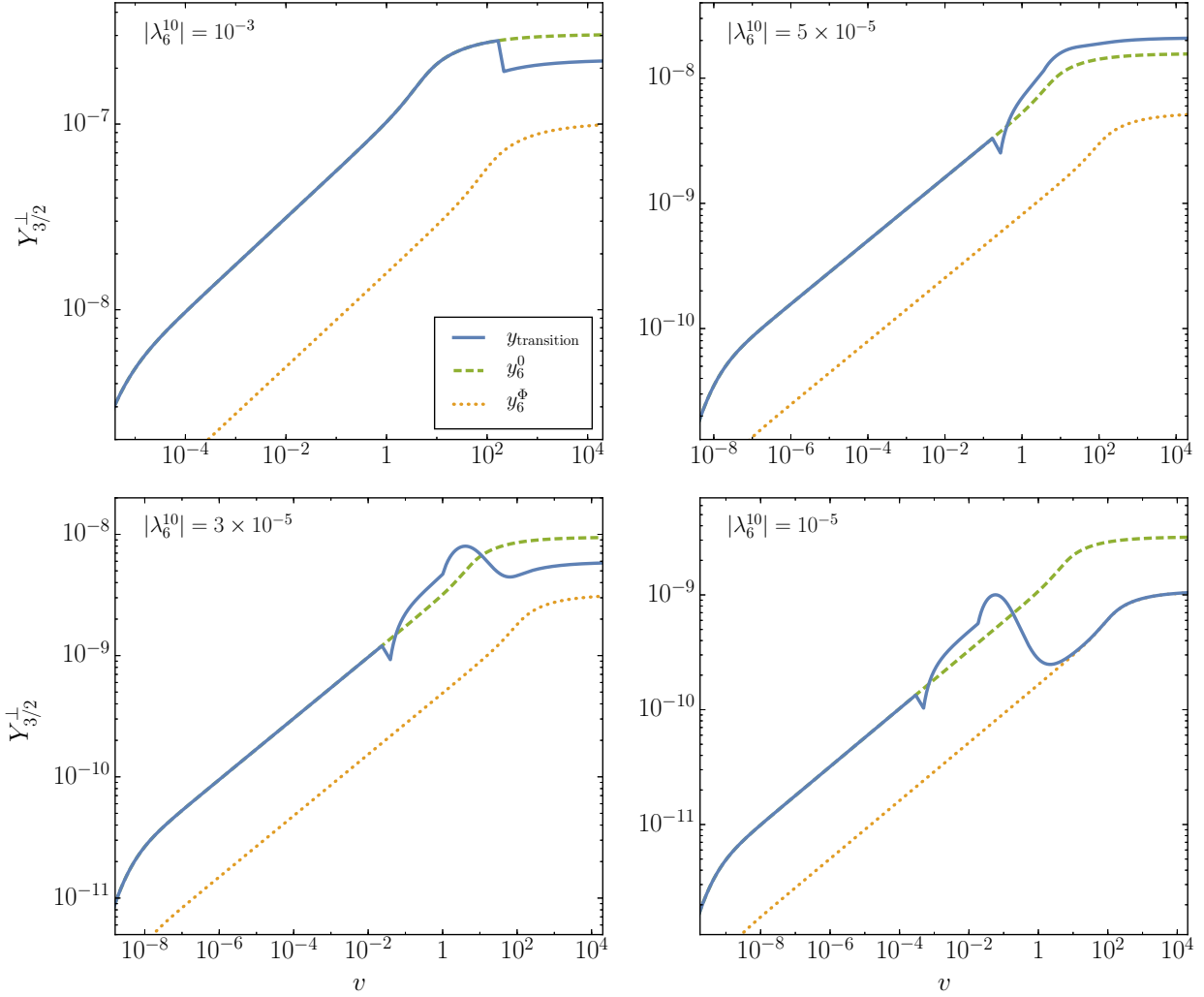}
    \caption{\em Time evolution of the transverse gravitino yield during and after reheating, for four different values of the Yukawa coupling $\lambda_6^{10}$, assuming strong reheating with $\Lambda_c=10\barM$. The continuous curves correspond to the numerical integration of (\ref{eq:n32eq}) accounting for the GUT phase transition. The dashed curves and the dotted curves disregard the occurrence of the transition, assuming instead a fixed Yukawa coupling $y=y_6^0$ and $y=y_6^\Phi$, respectively. }
    \label{fig:multi_grav}
\end{figure} 
As we have emphasized throughout Section~\ref{sec:2comp}, the no-scale flipped ${\rm SU}(5) \times {\rm U}(1)$ model that we study does not lead in general to a constant Yukawa coupling $y$ during reheating. This fact has important consequences for the production of gravitinos during inflaton decay. Let us for simplicity specialize to the strong reheating scenario discussed in Section~\ref{sec:strongreh}. Assume first that the Yukawa coupling $|\lambda_6^{10}|\gtrsim 2.7\times 10^{-5} $, that is $T_{\rm reh}\gtrsim 0.47\,\Lambda_c$ (see Eq. (\ref{eq:yminstr})). Since the reheating temperature is above the phase transition threshold, the Yukawa coupling $y=y_6^0$ is constant during reheating, and the gravitino yield would in principle be given by Eq.~(\ref{eq:y32}), with the dilution given by Eq. (\ref{strongD}). The upper left panel in Fig.~\ref{fig:multi_grav} shows the growth of the (transverse) gravitino yield~\footnote{The total gravitino yield can be obtain by multiplying the transverse yield by $(1+ 0.56 m_{1/2}^2/m_{3/2}^2)$.} during and after reheating for $|\lambda_6^{10}|=10^{-3}$ accounting for the phase transition (solid, blue), compared to that assuming $y=y_6^0$ (dashed, green) and $y=y_6^\Phi$ (dotted, orange). The end of reheating occurs in this case at $v\sim 1$. After reheating ends, around $v\sim166$, the number of light degrees of freedom in the plasma is reduced from 1545/4 to 915/4 in accordance with (\ref{eq:dofub}), and the GUT phase transition occurs at $v\sim 10^3$. Note that up until the end of reheating, the gravitino yield follows the $y_6^0$ prediction, growing as $Y_{3/2} \propto 1/T$~\cite{gmop}. When the aforementioned change in the number of degrees of freedom occurs, the yield is reduced with respect to the $y_6^0$ result. As is well known~\cite{rs,EGNOP,gmop,ag}, and also evident from the figure, the gravitino abundance is sensitive primarily to the temperature at the final stages of reheating, after the production of entropy has ceased. Given that in this scenario the phase transition is delayed with respect to the end of reheating, the reduction in $Y_{3/2}$ due to the change in $g$ is permanent, as can be seen for $v\gg 1$. It must be noted that, despite the
difference of the results accounting for the phase transition and assuming $y=y_6^0$, their late-time yields and CDM abundances would be similar, since at the end of production $g=915/4$ for the former, while $g = 1545/4$ for the later.

The upper right panel of Fig.~\ref{fig:multi_grav} demonstrates the evolution of the transverse yield for $|\lambda_6^{10}|= 5\times 10^{-5}$ in the strong reheating scenario with $\Lambda_c=10\barM$. In this case, the reduction in the number of degrees of freedom occurs around $v\sim 0.2$, shortly before the end of reheating, while the GUT phase transition takes place around $v\sim 4$. In contrast to the previous scenario, the change in degrees of freedom results in a yield which momentarily reduced with respect to the $y_6^0$ result, but it is shortly afterwards increased due to the enhanced production rate at lower temperature. As in the previous case, since the transition occurs during the latest stages of entropy production, the difference between the result accounting for the transition and assuming $y=y_6^0$ is permanent; in this case it leads to a net enhancement in $Y_{3/2}$.

The lower left panel of Fig.~\ref{fig:multi_grav} shows the time dependence of the transverse yield for $|\lambda_6^{10}|= 3\times 10^{-5}$, assuming strong reheating. Similarly to the previous case, the change in degrees of freedom around $v\sim 0.02$ leads to the enhancement of $Y_{3/2}$ relative to the $y=y_6^0$ curve. However, in this case the SU(5)$\times$ U(1) $\rightarrow$ SU(3)$\times$SU(2)$\times$U(1) transition occurs around $v\sim 1$, right by the end of reheating. At this moment the universe becomes radiation dominated, but the decay of the inflaton is not yet complete. Thus, although for $v \gtrsim 1$ the gravitino yield starts to freeze due to the absence of significant entropy production, it is posteriorly diluted around $v\sim 10$, when the $s\rightarrow \nu^c_1~\Phi$ decay channel dominates, which releases an additional amount of entropy. For the particular parameter values chosen herein, this entropy release overcompensates for the aforementioned enhancement, and results in a yield that is reduced relative to that assuming a constant $y=y_6^0$.

For $|\lambda_6^{10}|\lesssim 2.7\times 10^{-5} $, or equivalently $T_{\rm reh}\lesssim 0.47\,\Lambda_c$, the GUT phase transition will occur before the end of reheating. We therefore expect that in this case the final yield will be given by (\ref{eq:y32}), with the Yukawa coupling $y_6^\Phi$. The lower right panel of Fig.~\ref{fig:multi_grav} shows the evolution of the transverse gravitino yield for $Y_{3/2}$ with $|\lambda_6^{10}|=10^{-5}$. Note that before the phase transition occurs, for $v \lesssim 0.02$, the yield follows the $y=y_6^0$ curve. After the phase transition takes place, the gravitinos are diluted at the same rate as the $s$-decay products and $Y_{3/2}\sim\,{\rm const.}$, until the latter start being produced copiously again around $v\sim 0.1$, at which point $Y_{3/2}$ decreases until it reaches its equilibrium value with respect to the $\nu^c_1 \Phi$ channel, shortly after the end of reheating\footnote{The end of the reheating process in this case is dominated by $\nu^c_1~\Phi$ final states, which do not themselves thermalize as their interactions are all suppressed by the GUT scale. However, as we discuss in the next Section, when $\lambda_6^{10}$ is small, $\nu^c_1$ is relatively light and decays quickly. Its decay products thermalize rapidly allowing for the production of gravitinos as shown in Fig.~\ref{fig:multi_grav} for small $\lambda_6^{10}$.}. Note that in this case all evidence of the finite duration of the GUT phase transition has been erased.

\begin{figure}[!ht]
\centering
    \includegraphics[width=0.9\textwidth]{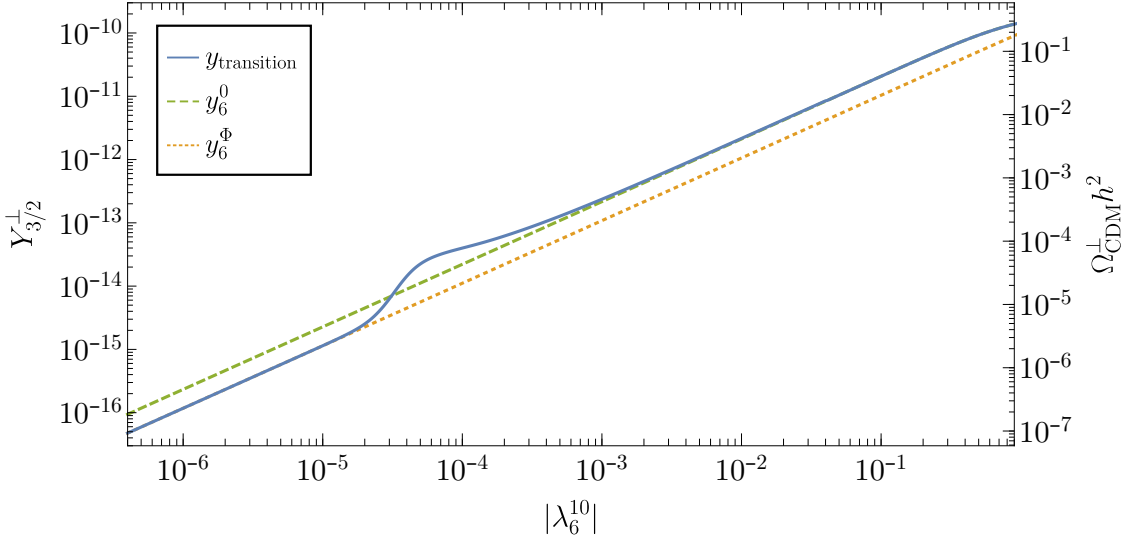}
    \caption{\it The transverse gravitino yield at $T\ll 1\,{\rm MeV}$ as a function of the Yukawa coupling $\lambda_6^{10}$ assuming strong reheating with $\Lambda_c=10\barM$. The left vertical axis corresponds to the numerically calculated yield including the dilution factor $\Delta$ given by (\ref{strongD}). The right vertical axis shows the corresponding CDM closure fraction, assuming $m_{\rm LSP}=100\,{\rm GeV}$. }
    \label{fig:yield_strong}
\end{figure}

The effect of the GUT phase transition on the relic gravitino yield, and hence on the primordial dark matter abundance, is shown in Fig.~\ref{fig:yield_strong} as a function of $\lambda_6^{10}$, for strong reheating with $\Lambda_c=10\barM$. For the curves labeled $y_{\rm transition}$ and $y_6^\Phi$ we have considered the dilution factor $g(T)/g_{\rm reh} \simeq 0.017$ due to the difference in the number of degrees of freedom between the MSSM and the Standard Model at $T\ll 1\,{\rm MeV}$; for the $y_6^0$ curve, the dilution factor is given by $g(T)/g_{\rm reh} \simeq 0.010$. We have also accounted for the entropy dilution factor $\Delta$ due to the late decay of the flaton $\Phi$, which in the strong reheating case is given by (\ref{strongD}); for definiteness we have considered $\lambda_{1,2,3,7}^{-2}\sim 1$. For $|\lambda_6^{10}|\gtrsim 2.7\times 10^{-5}$ the final yield has a numerical value close to that assuming that the $F_1 \bar{H}$ operates exclusively, although it is larger due to the difference in degrees of freedom discussed above. As expected, for larger values of the Yukawa coupling, the agreement between both is better. For $|\lambda_6^{10}|\lesssim 10^{-5}$, the resulting yield is indistinguishable from its value assuming the $\nu^c_1 \Phi$ channel operates exclusively. Note that in the absence of a significant contribution to the gravitino yield from the production of the longitudinal component, the observed CDM closure fraction is saturated at $|\lambda_6^{10}|\sim 0.3$. Nevertheless, if $m_{1/2}\gg m_{3/2}$ or $m_{\rm LSP} > 100$ GeV, the dark matter abundance may be easily saturated for much smaller values of the Yukawa coupling. We also note that for $|\lambda_6^{10}|\gtrsim 0.3$, in the edge of validity of the perturbative approximation, the numerically computed yield deviates from the power law dependence (\ref{eq:y32}), which is strictly valid only for $|y|\ll 1$~\cite{EGNOP}. 

\begin{figure}[!ht]
\centering
    \includegraphics[width=0.9\textwidth]{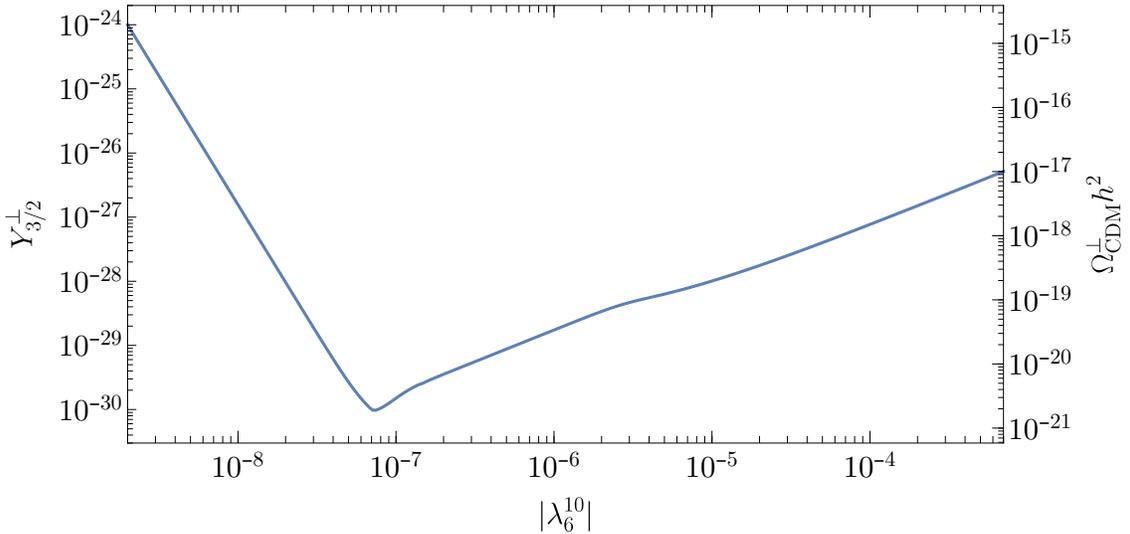}
    \caption{\it The transverse gravitino yield at $T\ll 1\,{\rm MeV}$ as a function of the Yukawa coupling $\lambda_6^{10}$ assuming moderate reheating with $\Lambda_c=0.4\barM$. The left vertical axis corresponds to the numerically calculated yield including the dilution factor $\Delta$ given by (\ref{eq:entropyv1}), (\ref{eq:entropyv2}) and (\ref{eq:entropyv3}). The right vertical axis shows the corresponding CDM closure fraction, assuming $m_{\rm LSP}=100\,{\rm GeV}$. }
    \label{fig:yield_moderate}
\end{figure}

Fig.~\ref{fig:yield_moderate} shows the relic gravitino yield and dark matter abundance as a function of $\lambda_6^{10}$, for moderate reheating with $\Lambda_c=0.4\barM$. The difference with Fig.~\ref{fig:yield_strong} is striking, and it is due to the dependence of the entropy dilution factor $\Delta$ on the $\lambda_6$ coupling, illustrated in Fig.~\ref{fig:entropy}. For a constant $\Delta$, the shape of the yield curve would be similar to that in the strong reheating scenario, centered at $|\lambda_6^{10}|=7\times 10^{-8}$, corresponding to $T_{\rm reh}\simeq 0.03\,\Lambda_c$.  However, due to the Yukawa dependence of $\Delta$, the gravitino yield decreases for $|\lambda_6^{10}|\lesssim 7\times 10^{-8}$, and increases for $|\lambda_6^{10}|\gtrsim 7\times 10^{-8}$.
 It is clear in this case that the enormous entropy dilution prevents the saturation of the observed dark matter abundance by its production through gravitino decays.\\

\section{CMB constraints}
\label{sec:cmb}

We now consider the constraints on the decay rate of the inflaton that result from the altered cosmological history between reheating and BBN. The entropy increase due to the decay of the flaton at late times modifies the relation between the decay rate and the number of $e$-folds to the end of inflation. Denoting by $k_*$ the comoving pivot scale, we can write the relative size of its present physical wavelength to the present horizon as follows,
\beq
\frac{k_*}{a_0 H_0} \;=\; \frac{a_* H_*}{a_0 H_0} \;=\; e^{-N_*} \frac{H_*}{\rho_{\rm end}^{1/4}} \frac{a_{\rm end}}{a_{\rm reh}}\left(\frac{\rho_{\rm end}}{\rho_{\rm reh}}\right)^{1/4}  \frac{a_{\rm reh}}{a_{d\Phi}}\left(\frac{\rho_{\rm reh}}{\rho_{ d\Phi}}\right)^{1/4}  \left(\frac{a_{d\Phi} \rho_{d\Phi}^{1/4}}{a_0H_0}\right)\,,
\eeq
where $a_0$ and $H_0$ denote the present cosmological scale factor and Hubble expansion rate, respectively, $N_*$ is the number of $e$-folds to the end of inflation, $H_*$ is the expansion rate at horizon crossing, and where the subindices ``end'', ``reh'' and $d\Phi$ indicate evaluation of the corresponding energy densities $\rho$ and scale factors $a$ at the end of inflation, the end of reheating and the decay of $\Phi$, respectively. Solving for $N_*$ we obtain the following expression
\beq\label{eq:Nstar}
N_* \;=\; -\ln\left(\frac{k_*}{a_0H_0}\right) + \ln\left(\frac{H_*}{\rho_{\rm end}^{1/4}}\right) + \frac{1}{4}\ln\left( \frac{\rho_{\rm end} a_{\rm end}^4 }{\rho_{\rm reh} a_{\rm reh}^4}\right) + \ln\left(\frac{a_{d\Phi} \rho_{d\Phi}^{1/4}}{a_0H_0}\right)+ \frac{1}{4}\ln\left(  \frac{\rho_{\rm reh} a_{\rm reh}^4}{\rho_{ d\Phi} a_{d\Phi}^4 }\right)\,.
\eeq
This result differs from the standard relation~\cite{planck15,Aghanim:2018eyx,planck18,LiddleLeach,MRcmb,egnno2} in that the last term explicitly accounts for the increase in entropy
\cite{Easther:2013nga,Adshead:2010mc} between the end of inflation and today due to the decay of the flaton $\Phi$,
\begin{align}
 \frac{1}{4}\ln\left(  \frac{\rho_{\rm reh} a_{\rm reh}^4}{\rho_{ d\Phi} a_{d\Phi}^4 }\right) \;&=\; \frac{1}{3} \ln\left(\frac{s_{\rm reh} a_{\rm reh}^3}{s_{d\Phi} a_{d\Phi}^3}\right) + \frac{1}{12}\ln\left(\frac{g_{d\Phi}}{g_{\rm reh}}\right) \nonumber \\
 & = -\frac{1}{3}\ln\Delta + \frac{1}{12}\ln\left(\frac{g_{d\Phi}}{g_{\rm reh}}\right)\,.
\end{align}
Evaluating the right-hand side of (\ref{eq:Nstar}) at the Planck pivot point $k_* = 0.05/{\rm Mpc}$, corresponding to $k_*/a_0H_0 = 221$, we obtain
\begin{align} 
N_* \;&=\; 62.04 + \ln\left(\frac{H_*}{\rho_{\rm end}^{1/4}}\right) + \frac{1}{4}\ln\left( \frac{\rho_{\rm end} a_{\rm end}^4 }{\rho_{\rm reh} a_{\rm reh}^4}\right) - \frac{1}{12}\ln g_{\rm reh}  - \frac{1}{3}\ln\Delta \nonumber \\
&=\; N_*^{\rm STH} - \frac{1}{3}\ln\Delta\,,
\label{eq:Nstar2}
\end{align}
where STH denotes a standard thermal history, with entropy conservation following the end of reheating. 

Let us first evaluate (\ref{eq:Nstar2}) assuming the strong reheating conditions are verified; when this is the case the entropy dilution factor has the $\lambda_6$-independent value (\ref{strongD}). Although an analytical approximation for $N_*$ is available assuming a pure Starobinsky potential for the inflaton $s$~\cite{EGNO5}, we will solve (\ref{eq:Nstar2}) numerically to allow for added generality the potential
\beq
V(s) \;=\; \frac{3}{4}m_s^2 M_P^2 \left(1-e^{-\sqrt{2/3}\,s/M_P}\right)^2 + 81 \zeta m_s M_P^3 \sinh^4(s/\sqrt{6} M_P)\left(\tanh(s/\sqrt{6}M_P)-1\right)\,,
\eeq
where
\beq
\zeta \;=\; \sum_i (M_P/\mu_{ii})(\lambda_8^{00i})^2 + {\rm h.c.}\,,
\eeq
which arises after integrating out the dynamics of the heavy singlets $\phi_i$ during inflation, assuming a strongly segregated inflaton sector, $\lambda_8^{0ij}\lesssim \mu^{ij}/M_P$. This strong segregation condition and the bound $\lambda_8^{00i}\lesssim 10^{-5}$ are sufficient to ensure the adiabatic evolution of the heavy singlets during reheating, when the energy density during reheating is dominated by the oscillating inflaton. For $\lambda_8^{0ij}\gtrsim \mu^{ij}/M_P$ and/or multiple light $\phi_i$ adiabaticity is in general badly violated and the simple picture of reheating that we have considered in this paper breaks down~\cite{egnno2}.

\begin{figure}[!t]
\centering
    \includegraphics[width=0.9\textwidth]{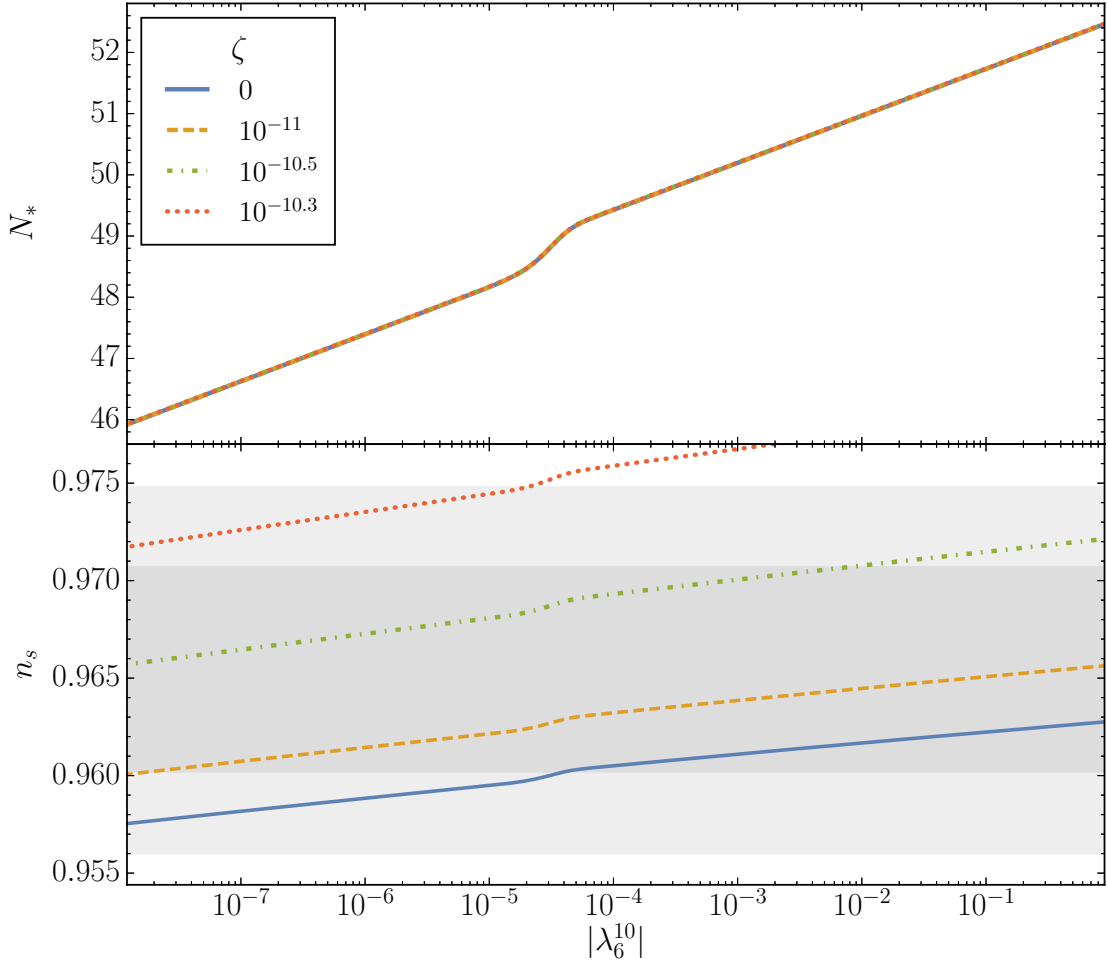}
    \caption{\it The number of $e$-folds to the end of inflation (upper panel) and the scalar tilt (lower panel) as functions of the Yukawa coupling $\lambda_6^{10}$ and the deformation parameter $\zeta$ assuming strong reheating with $\Lambda_c=10\barM$. The light gray (gray) shaded area corresponds to the 95\% (68\%) PBK CL region at low tensor-to-scalar ratio~\cite{Ade:2018gkx}. }
    \label{fig:cmb_strong_lambda}
\end{figure}
Figure~\ref{fig:cmb_strong_lambda} shows the result of the numerical solution of (\ref{eq:Nstar2}) for a few values of $\zeta\ll 1$, with $\Lambda_c=10\barM$ and $\lambda_{1,2,3,7}^{-2}\sim 1$. The upper panel shows the dependence of $N_*$ on $\lambda_6^{10}$. Following our discussion of the gravitino abundance, for $|\lambda_6^{10}|\gtrsim 7\times 10^{-5}$ the number of $e$-folds follows the Starobinsky prediction with the Yukawa coupling given by $y_6^0$, corresponding to the $s\rightarrow F_1 \bar{H}$ decay channel. For $|\lambda_6^{10}|\lesssim 10^{-5}$, $N_*$ is given by the Starobinsky result assuming the $\nu^c_1 \Phi$ channel dominates. Note that $N_*$ is insensitive to the degree of deformation parametrized by $\zeta$ for the range of values considered. The lower panel shows the dependence of the scalar tilt $n_s$ on $\lambda_6^{10}$; shaded in gray are the Planck+BICEP2/Keck (PBK) 68\% and 95\% confidence level regions for $n_s$ at low tensor-to-scalar ratio~\cite{Ade:2018gkx}. The effect of the potential deformation is clear in this case, where an increasing $\zeta$ leads to a steeper inflaton potential and therefore to an increased $n_s$ relative to $\zeta=0$. For pure Starobinsky, the scalar tilt lies within the 95\% (68\%) CL region for $|\lambda_6^{10}|\gtrsim 6\times 10^{-11}$ ($4\times 10^{-5}$). 

\begin{figure}[!t]
\centering
    \includegraphics[width=0.9\textwidth]{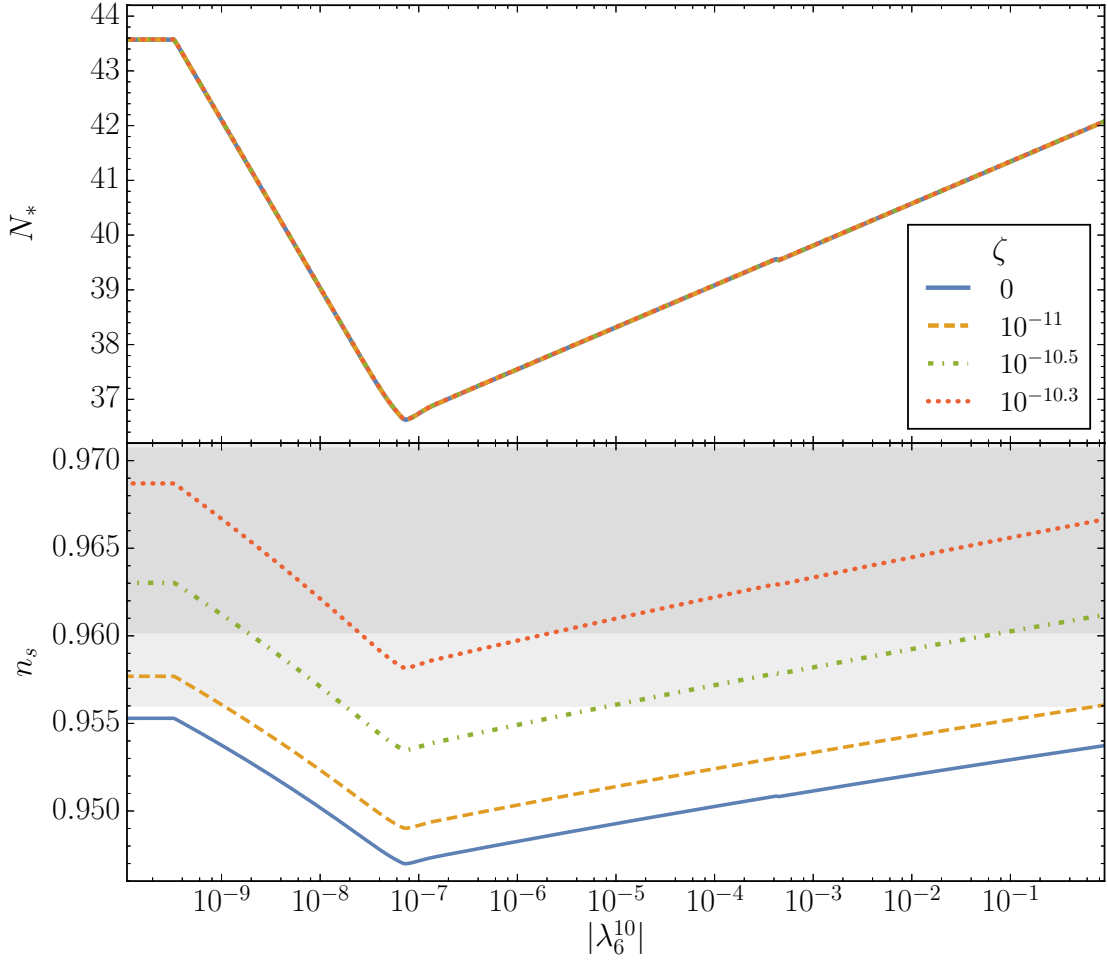}
    \caption{\it The number of $e$-folds to the end of inflation (upper panel) and the scalar tilt (lower panel) as functions of the Yukawa coupling $\lambda_6^{10}$ and the deformation parameter $\zeta$ assuming moderate reheating with $\Lambda_c=0.4\barM$. The light gray (gray) shaded area corresponds to the 95\% (68\%) PBK CL region at low tensor-to-scalar ratio~\cite{Ade:2018gkx}. }
    \label{fig:cmb_moderate_lambda}
\end{figure}
Let us now consider moderate reheating, with dilution factor $\Delta$ given by (\ref{eq:entropyv1}), (\ref{eq:entropyv2}) and (\ref{eq:entropyv3}). Figure~\ref{fig:cmb_moderate_lambda} displays the solution of (\ref{eq:Nstar2}) for the same values of $\zeta$ as Fig.~\ref{fig:cmb_strong_lambda}, assuming $\Lambda_c=0.4\barM$ and $\lambda_{1,2,3,7}^{-2}\sim 1$. Analogously to the strong reheating case, both panels closely mimic the gravitino abundance curve shown in Fig.~\ref{fig:yield_moderate}: $N_*$ and $n_s$ are approximately constant for $|\lambda_6^{10}|\lesssim 3\times 10^{-10}$, and are decreasing functions for $3\times 10^{-10}\lesssim |\lambda_6^{10}|\lesssim 7\times 10^{-8}$, and increasing for $|\lambda_6^{10}|\gtrsim 7\times 10^{-8}$. As can be appreciated in the upper panel, for all values of $\zeta$ considered the functional dependence of $N_*$ on the Yukawa coupling is seemingly identical. The lower panel shows the effect of the potential deformation on the scalar tilt. It is clear how the large release of entropy due to the decay of the coherent flaton has pushed the pure Starobinsky result outside the 95\% Planck-BK CL region, which implies that the model would only be consistent with observations for $\zeta\gtrsim 10^{-10.3}$.\\

\section{Neutrino Mass Structure and Leptogenesis} \label{sec:neutrinos}

As we have seen above, the coupling $\lambda_6^{i 0}$ plays a
crucial role in both reheating and the generation of gravitinos. In this
Section, we discuss a third role of this coupling---the generation of
light neutrino masses. For clarity, we study here  a
single-generation version of the neutrino mass matrix for $\nu_{i}$,
$\nu^c_{i}$, and $\tilde{S}$ (the fermionic partner of the
inflaton $S$), as in Ref.~\cite{egnno2}:
\begin{equation}
 {\cal L}^{(i)}_{\rm mass}=
-\frac{1}{2}
\left(
\begin{matrix}
{\nu}_{i} & {\nu}_{i}^c & \tilde{S}
\end{matrix}
\right) 
\begin{pmatrix}
 0 & \lambda_2^{i i}\langle \bar{h}_0\rangle & 0 \\
\lambda_2^{i i}\langle \bar{h}_0\rangle & 0 &
 \lambda_6^{i 0}\langle \tilde{\nu}_{\bar{H}}^c\rangle \\
0 & \lambda_6^{i 0}\langle \tilde{\nu}_{\bar{H}}^c\rangle & m_s
\end{pmatrix}
 \left(
\begin{matrix}
\nu_{i} \\ \nu_{i}^c \\ \tilde{S}
\end{matrix}
\right)
+ {\rm h.c.}~,
\label{massm}
\end{equation}
neglecting for simplicity mixing with the other generations as well as potential CP
phases in the couplings in the mass matrix. (For more
generic expressions, see Ref.~\cite{Ellis:1993ks}.) The mass eigenvalue
of the lightest state, which corresponds to one of the electroweakly-interacting (active) neutrinos,
is then given by
\begin{equation}
 m_{\nu_{i}} \simeq \frac{m_s\left(\lambda_2^{ii}\langle 
 \bar{h}_0\rangle\right)^2}{\left(\lambda_6^{i0} \langle 
 \tilde{\nu}^c_{\bar{H}} \rangle \right)^2} 
\simeq 
\frac{m_s m_{u_i}^2}{\left(\lambda_6^{i0} 
\langle \tilde{\nu}^c_{\bar{H}} \rangle \right)^2}
~,
\label{eq:linumassesb}
\end{equation}
where we have used in the second part of the
equation the relation that follows from Yukawa unification
in flipped SU(5)$\times$U(1).
The mass eigenstates of the heavier neutrinos were already given in Eq. (\ref{hnumass}). The inflaton mass $m_s$ is fixed to be $\simeq 3 \times
10^{13}$~GeV, while $\langle
\tilde{\nu}^c_{\bar{H}} \rangle$ is a GUT-scale expectation value,
which is fixed at $10^{16}$~GeV. We thus
find from Eq.~\eqref{eq:linumassesb} that one of the light neutrino
masses, $m_{\nu_{i}}$, is predicted as a function of the coupling
$\lambda_6^{i 0}$. 

The masses of the two remaining families (with index $j$) 
are obtained from a 
mass matrix similar to Eq.~(\ref{massm}) 
with the replacement of $\lambda_2^{ii}$ with $\lambda_2^{jj}$,  $\lambda_6^{i0}$ with $\lambda_6^{ja}$, $m_s$ with $\mu^{ab} > m_s$ where
$a,b \ne a'$, i.e., the singlets involved in the neutrino mass matrix are 
orthogonal to the singlet generating the $\mu$ term through $\lambda_7^{a'}$ and we have assumed that $\lambda_6^{i a'} = 0$.  Recall
that $a'$ corresponds to the linear combination of $\phi_a$ such that
$\langle \phi_{a'} \rangle \ne 0$. See \cite{egnno2} for further details.

\begin{figure}[t]
\centering
\includegraphics[height=70mm]{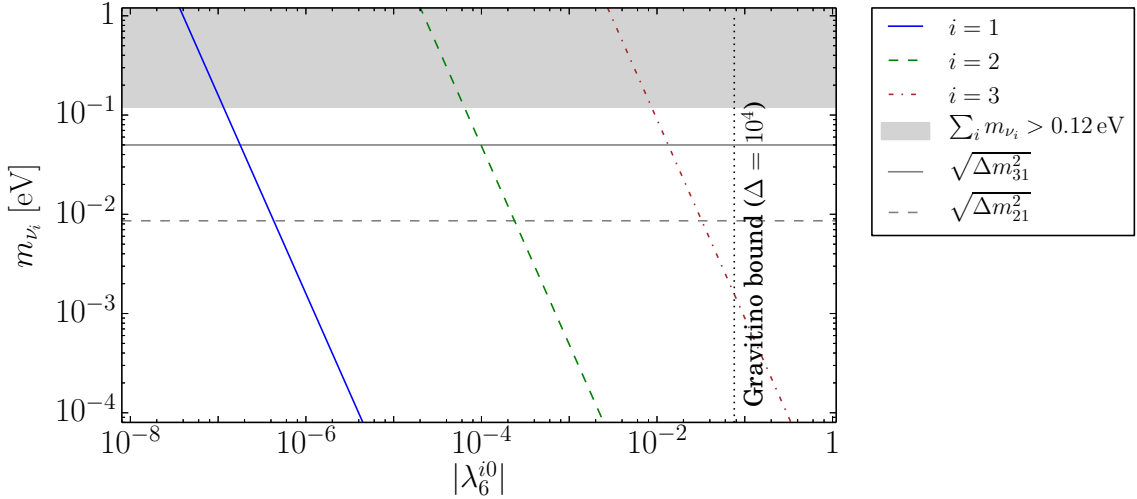}
\caption{\it 
Neutrino masses $m_{\nu_i}$ as
functions of the coupling $|\lambda_6^{i 0}|$ for $i =
1,2,3$ as the blue solid, green dashed, and brown dashed-dotted lines,
respectively. We set $m = 3 \times 10^{13}$~GeV and $\langle
\tilde{\nu}^c_{\bar{H}} \rangle = 10^{16}$~GeV. 
The gray shaded area is excluded by the limit on the sum
of the neutrino masses set by the Planck 2018 data
\cite{Aghanim:2018eyx}: $\sum_i m_{\nu_i} < 0.12$~eV. 
The vertical black dotted line corresponds to the limit coming from the gravitino overproduction for $\Delta = 10^4$ and $m_{\text{LSP}} = 100$~GeV. 
We also show the values of $\sqrt{\Delta m_{31}^2}$
 and $\sqrt{\Delta m_{21}^2}$, which are taken from the latest global
 fit \cite{nufit, Esteban:2016qun}, as the solid and dashed gray lines,
 respectively. }
  \label{fig:mnu}
\end{figure}

In Fig.~\ref{fig:mnu} we plot neutrino masses $m_{\nu_i}$ as
functions of the coupling $|\lambda_6^{i 0}|$ for $i =
1,2,3$ as the blue solid, green dashed, and brown dashed-dotted lines,
respectively, for $m_s = 3 \times 10^{13}$~GeV and $\langle
\tilde{\nu}^c_{\bar{H}} \rangle = 10^{16}$~GeV. 
The gray shaded area is excluded by the limit on the sum
of the neutrino masses set by the Planck 2018 data
\cite{Aghanim:2018eyx}: $\sum_i m_{\nu_i} < 0.12$~eV. 
As we see, the Planck bound gives lower limits on the $\lambda_6^{i 0}$:
\begin{equation}
|\lambda_6^{1 0}| \gtrsim 10^{-7}, \; |\lambda_6^{2 0}| \gtrsim 10^{-4}, \; 
|\lambda_6^{3 0}|\gtrsim 10^{-2} \, .
\end{equation}
We then conclude from the limit on $\lambda_6^{1 0}$ and comparing with Eqs.~\eqref{eq:yminstr} and \eqref{eq:l6truemod} that only strong (moderate) reheating is possible when
$\Lambda_c \gtrsim 2.3 \barM$ ($\Lambda_c \lesssim 2.3
\barM$); in other words, the weak reheating
scenario is incompatible with the Planck limit on $\sum_i m_{\nu_i}$. Moreover, in the
case of the moderate reheating scenario, the entropy release is 
maximum, $\Delta \sim 5 \times 10^{17}$, as shown in Eq.~\eqref{eq:entropyv1} and Fig.~\ref{fig:entropy}.

We can also obtain a lower bound on the neutrino mass from the upper 
bound on $\lambda_6^{i 0}$.  We recall that, since $\lambda_6^{i 0}$ controls directly the
reheating temperature and therefore the gravitino abundance, we obtained
in Eq.~(\ref{ocdm}) the relic abundance of cold dark matter produced
in gravitino decays. To avoid the overabundance of dark matter,
we have an upper limit on $\lambda_6^{i 0}$ from Eq.~(\ref{ylimit})
\beq
|\lambda_6^{i 0}| < 7.6\times 10^{-2} \times \biggl(\frac{ \Delta}{10^{4}} \biggr) \left( \frac{100 {\rm GeV}}{m_{\rm LSP}} \right)
\left(1+0.56 \frac{m_{1/2}^2}{m_{3/2}^2}\right)^{-1} \, ,
\eeq
shown as the vertical dotted line in Fig. \ref{fig:mnu}.
Thus we obtain
\beq
m_{\nu_i} >  \biggl(\frac{ \Delta}{10^{4}} \biggr)^{-2} \left( \frac{m_{\rm LSP}}{100 {\rm GeV}} \right)^2  \left(1+0.56 \frac{m_{1/2}^2}{m_{3/2}^2}\right)^{2}
\times
\begin{cases}
2.8 \times 10^{-13}~\text{eV} & (i = 1) \\
8.5 \times 10^{-8}~\text{eV} & (i = 2)\\
1.6 \times 10^{-3}~\text{eV} & (i = 3)
\end{cases}
\, ,
\eeq
using our canonical choices for $m_s$ and $\langle
\tilde{\nu}^c_{\bar{H}} \rangle$ and $m_{u,c,t} = 2.3 \times 10^{-3}, 1.27, 174.2$ GeV.
As we see, all of the three cases are compatible with neutrino data. We also note in passing that the LSP abundance from the non-thermal gravitino decay agrees to the observed dark matter density at the border of the gravitino bound ($|\lambda_6^{i 0}| \simeq 7.6\times 10^{-2}$), if we assume the thermal relic abundance of the LSP is negligibly small.

\begin{figure}[ht]
\centering
\includegraphics[height=70mm]{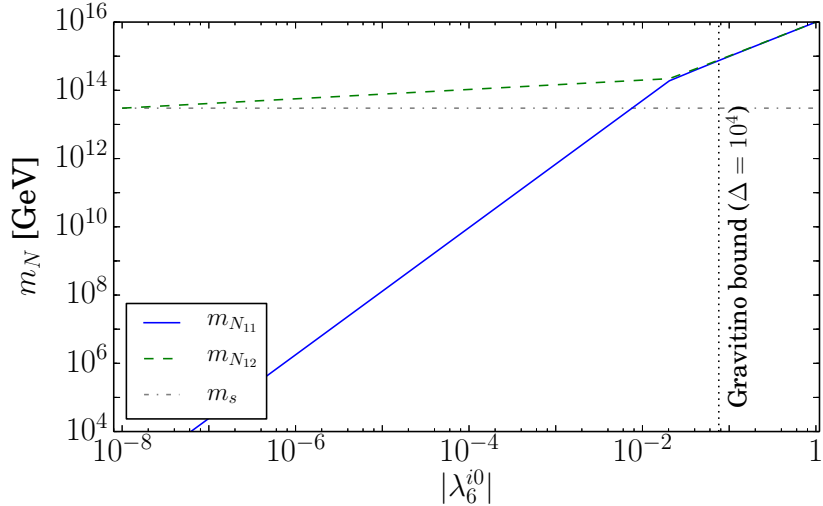}
\caption{\it 
The masses of the heavier neutrino states as functions of 
$\lambda_6^{i 0}$. The horizontal gray dotted line 
shows $m_s = 3 \times 10^{13}$~GeV. 
The vertical black dotted line corresponds to the limit coming from the gravitino overproduction for $\Delta = 10^4$ and $m_{\text{LSP}} = 100$~GeV. 
}  
  \label{fig:mn}
\end{figure}

In Fig.~\ref{fig:mn} we show the masses of the heavier neutrino states as functions of $|\lambda_6^{i 0}|$ together with the inflaton mass $m_s = 3 \times 10^{13}$~GeV, shown by the horizontal gray dotted line. As can be seen from the figure, one of the heavier neutrino states is lighter than the inflaton mass for $|\lambda_6^{i 0}| \lesssim 4 \times 10^{-3}$; in this case, the inflaton can always decay into flaton and the heavy neutrino state as we discussed in the previous sections. 

\begin{figure}[ht]
\centering
\includegraphics[height=70mm]{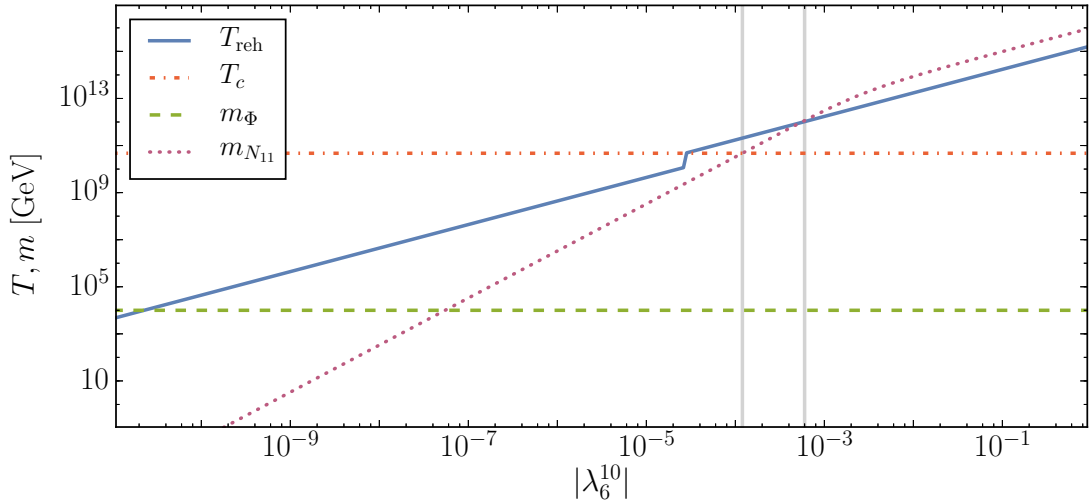}
\caption{\it 
The masses of the right-handed neutrino and flaton as well as the reheat and GUT transition temperatures as functions of 
$\lambda_6^{1 0}$.  
The vertical grey lines correspond to the limits where  $m_{\nu^c_1} = T_{c}$
(left line) and $m_{\nu^c_1} = T_{\rm reh}$ (right line). 
}  
  \label{fig:mnTr}
\end{figure}

Our results also impact the generation of a baryon asymmetry through leptogenesis \cite{fy}. As we have just noted, in the case of the moderate reheating
scenario, a huge amount of entropy is released, which makes it impossible to explain the observed baryon asymmetry with leptogenesis. We thus focus on the strong reheating scenario. 
In this case, when $|\lambda_6^{i 0}| > 2.7 \times 10^{-5}$ the GUT symmetry is not broken until reheating
is complete, and thus right-handed (singlet) neutrinos are initially massless. They are copiously produced
and in thermal equilibrium with their number density given by
\begin{equation}
 \frac{n_{\nu^c_{i}}}{s} \; \simeq \; \frac{135\zeta (3)}{4\pi^4 g_{*S}} ~,
 \label{eq:nnuovs}
\end{equation}
where $s$ denotes the entropy density. When the temperature of the
Universe becomes $\lesssim 0.47 \Lambda_c$, the GUT phase transition occurs and these
right-handed neutrinos acquire large masses. Since the phase transition proceeds incoherently, and its time scale is quite short, we can assume that right-handed neutrinos become massive almost instantaneously. These right-handed neutrinos are out of equilibrium after the phase transition and decay non-thermally \cite{NOT} if $\lambda_6^{i0} \gtrsim 10^{-4}$ to ensure $m_{\nu^c_i} > T_{c}$, where the transition temperature is $T_c = 0.47 \Lambda_c$ (see the left vertical line in Fig.~\ref{fig:mnTr}). 
In Fig. \ref{fig:mnTr}, we show the mass of the right-handed neutrino (as well as the flaton mass) and both the reheating and transition temperatures as functions of the coupling $\lambda_6^{10}$.
The
resultant baryon asymmetry generated through the non-thermal decay of
right-handed neutrinos is then estimated to be
\begin{align}
 \left|\frac{n_{\rm B}}{s}\right| &\simeq 
\bigl|\sum_{i} \epsilon_i\bigr| \times \biggl(\frac{28}{79}\biggr)
\biggl(
 \frac{135\zeta (3)}{4\pi^4 g_{*S} \Delta}
\biggr) 
\nonumber \\
&\simeq 
4\times 10^{-8} \times
\bigl|\sum_{i} \epsilon_i\bigr| \times
\biggl(\frac{g_{*S}}{1545/4}\biggr)^{-1}
\biggl(\frac{\Delta}{10^4}\biggr)^{-1} ~,
\label{eq:nbovs} 
\end{align}
where the CP violation is given by, {\it e.g.}, for $i = 1$  \cite{Ellis:1993ks, Covi:1996wh} 
\begin{equation}
 \epsilon_1 \simeq -\frac{3}{2\pi}\frac{1}
{\left(U_{\nu^c}^\dagger (\lambda_2^D)^2 U_{\nu^c}\right)_{11}}
 \sum_{j=2,3} {\rm Im} \left[\left(U_{\nu^c}^\dagger (\lambda_2^D)^2 U_{\nu^c}\right)^2_{j1}\right] 
\frac{m_{N_{11}}}{m_{N_{j1}}} ~,
\label{eq:epsilon}
\end{equation} 
where we have assumed $m_{N_{11}}\ll m_{N_{j1}}$
and $U_{\nu^c}$ is a mixing matrix associated with the diagonalization of
$\tilde{\nu}_i^c$ and $S$ in the basis where $\lambda_2$ is
diagonalized to $\lambda_2^D$ \cite{Ellis:1993ks}.
For $i = 2, 3$ (or for $m_{N_{11}}\simeq m_{N_{j1}}$), $\epsilon_i$ is obtained by replacing 1 with $i$ in Eq.~\eqref{eq:epsilon} and  $m_{N_{11}}/ m_{N_{j1}}$ by the mass function given in Ref.~\cite{Covi:1996wh}.
We note that all of the three right-handed neutrinos participate in
generating the lepton asymmetry. As one can see, the observed baryon asymmetry
($n_{\text{B}}/s \simeq 10^{-10}$) is reproduced when $|\sum_{i}
\epsilon_i| \simeq 3 \times 10^{-3}$, which is achievable as
can be seen from Eq.~(116) in Ref.~\cite{egnno2}. 

When the GUT transition occurs before reheating,
one of the conditions for the out-of-equilibrium decay of the right-handed neutrinos is $m_{\nu^c_1} > T_{\rm reh}$
(shown by the right vertical line in Fig. \ref{fig:mnTr}).
 As discussed 
in \cite{Ellis:1993ks}, due to mixing, the decay of the lightest right-handed neutrino
will be proportional to $|{\lambda_2^{33}}|^2 m_{\nu^c_1}$, where $\lambda_2^{33} \sim 1$
is representive of the top quark Yukawa coupling. Thus the right-handed neutrino is kept in thermal equilibrium by its decays and
inverse decays unless $m_{\nu^c_1} > T_{\rm reh}$,
cutting off the inverse decays.
Unless this condition is met, any asymmetry 
will be washed away and an asymmetry must be
then be produced at $T < m_{\nu^c_1}$.
However, as one can see from the figure,
when $T_c > T_{\rm reh}$, $m_{\nu^c_1} < T_{\rm reh}$.
We do not consider this case further.

In summary, by taking account of neutrino masses and leptogenesis, we find
\begin{itemize}
    \item Weak reheating is incompatible with the Planck limit on the sum of neutrino masses.
    \item The observed baryon asymmetry cannot be reproduced in moderate reheating.
    \item Both neutrino masses and the baryon asymmetry can be explained in strong reheating.
\end{itemize}
Moreover, if $|\lambda_6^{i 0}| \simeq 7.6\times 10^{-2}$ and the thermal relic abundance of the LSP is negligibly small, the non-thermal decay of the gravitino can explain the observed dark matter density of the Universe. This specific situation is quite intriguing as the single parameter $|\lambda_6^{i 0}|$ determines the reheating temperature, the lightest neutrino mass, and the dark matter relic abundance simultaneously, while the baryon asymmetry of the Universe can be explained by a modified version of thermal leptogenesis accompanied with a phase transition. We will investigate this specific scenario on another occasion \cite{EGNNO4}.

\section{Summary}
\label{sec:sum}

We have developed in this paper a prototypical scenario for physics from the string scale to that of neutrino masses.
Any such scenario should include mechanisms for inflation and the breaking of GUT and electroweak symmetries, as well
as mechanisms for neutrino masses (presumably via a seesaw) and baryogenesis (presumably via leptogenesis). The
scenario we have developed here is based upon elements that we have advocated in earlier papers, which we have
combined here in a consistent framework for the dynamics of the early universe. Although this dynamics is complicated,
we have found a successful realization of our scenario that is insensitive to most of the model parameters.

The general framework is that of no-scale supergravity, which has the virtues of being derivable as the effective
field theory of supersymmetric compactifications of string theory, on the one hand, and avoiding anti-de Sitter `holes'
in the effective potential and the `$\eta$ problem', on the other hand. Moreover, as we have emphasized previously,
it provides for a simple realization of cosmological inflation whose predictions resemble those of the Starobinsky
model \cite{Staro}, which are compatible with all the CMB measurements to date~\footnote{A general classification of
Starobinsky-like no-scale supergravity models has been given recently~\cite{ENOV}, but we have restricted our attention
here to the original realization proposed in~\cite{ENO6}.}. As we have discussed in Section~6, our
scenario yields values of the tensor-to-scalar ratio $r$ that are very similar to those of the Starobinsky model
and favours values of the scalar tilt $n_s$ that are consistent with current constraints. However, as seen in
Figs.~\ref{fig:cmb_strong_lambda} and \ref{fig:cmb_moderate_lambda}, our scenario offers the possibility of
constraining model parameters via future measurements of $n_s$.

Our scenario adopts the SU(5)$\times$U(1) flipped model of grand unification
summarized in Section~2, which has also been derived
from string theory~\cite{AEHN,cehnt}. 
This model avoids rapid baryon decay via dimension-5 operators, which are the bane of
other supersymmetric GUT models, and contains a suitable seesaw mechanism for obtaining small masses
for the active neutrinos, as seen in Eqs.~(\ref{massm}) and (\ref{eq:linumassesb}). We recall that in this model, the first stage of GUT symmetry
breaking is due to vev's for SU(3)$\times$SU(2)$\times$U(1)-singlet `heavy neutrino'-like fields in 
$\mathbf{10}$ and $\mathbf{\overline {10}}$ representations of SU(5), denoted by $H$ and $\bar H$. As seen
in Fig.~\ref{fig:mnu}, this mechanism yields masses for the light active neutrinos $m_{\nu_i}$ that are comfortably
consistent with the upper limit on $\sum_i m_{\nu_i}$ from the Planck 2018 data~\cite{Aghanim:2018eyx}.
As also discussed in Section~\ref{sec:neutrinos}, our scenario also provides for baryogenesis via
leptogenesis, with all three heavy neutrinos participating. 

However, these conclusions depend on the strength of reheating during the expansion of the universe, for
which we distinguish three scenarios that are classified in Fig.~\ref{fig:classification}. Indeed, we find
that weak reheating is incompatible with the Planck limit on the sum of neutrino masses, whereas
the observed baryon asymmetry cannot be reproduced in moderate reheating. However,
both neutrino masses and the baryon asymmetry can be explained in strong reheating, which we therefore prefer.

Our analysis of dynamics during the early universe and reheating has been set out in Section~4. As discussed
there, there are two stages of reheating in our scenario, one associated with the decay of the inflaton, which
is some combination of singlet fields $\phi_a$, and another associated with the GUT SU(5)$\times$U(1) $\to$
SU(3)$\times$SU(2)$\times$U(1) phase transition. The latter is associated with the flaton, a combination of
SU(3)$\times$SU(2)$\times$U(1) singlet fields in the $H$ and $\bar H$ multiplets, which evolve along a $D$-flat
direction, as described in Section~3. The conditions for the preferred possibility of strong reheating are discussed
in detail in Section~4.2.

One of the constraints on reheating scenarios is provided by gravitino production in the early universe,
which was discussed in Section~5. As is well known, there are constraints on the cosmological
gravitino abundance that are imposed, in particular, by the density of dark matter particles produced in its
decays~\footnote{We recall that $R$-parity is not strictly conserved in our flipped SU(5) model, and hence
the lightest supersymmetric particle (LSP) is not absolutely stable. However, as already mentioned,
the breaking of $R$-parity is sufficiently sequestered that the LSP lifetime is much longer than the 
age of the universe, so that it remains as good a cold dark matter candidate as in $R$-conserving models.}.

Many of the important aspects of our scenario are controlled by one key parameter, the coupling $\lambda_6$
between $\mathbf{10}$ matter, GUT Higgs and singlet fields: $\lambda_6 F {\bar H} \phi$ in (\ref{Wgen}).
This coupling controls inflaton decay - see Eq.~(24), the CMB observable $n_s$ - see Fig.~\ref{fig:cmb_strong_lambda},
and neutrino masses - see Eqs.~(\ref{massm}) and (\ref{eq:linumassesb}). It is non-trivial that acceptable values of all these quantities can be
obtained with a common value of $\lambda_6$, and it is interesting that measurements of $n_s$ and neutrino
masses could in principle be used to constrain better this parameter in the future.

As we have emphasized at the beginning of this Section, the ambitious scenario we have presented in this paper
is a prototype that is vulnerable to modification or exclusion by future data. Nevertheless, we hope and expect
that some of the considerations we have assembled and used in this paper may be incorporated usefully into
the future refined phenomenology of particle physics and cosmology below the string scale.

\section*{Acknowledgements}
The work of J.E. was supported partly by the United Kingdom STFC Grant ST/P000258/1 
and partly by the Estonian Research Council via a Mobilitas Pluss grant. The work of M.A.G.G. was supported by the DOE grant DE-SC0018216. The work of N.N. was supported by the Grant-in-Aid for Young Scientists B (No.17K14270) and Innovative Areas (No.18H05542). The work of D.V.N. was supported partly by the DOE grant DE-FG02-13ER42020 
and partly by the Alexander S. Onassis Public Benefit Foundation. The work of K.A.O. was supported partly
by the DOE grant DE-SC0011842 at the University of Minnesota.

\end{document}